\documentclass[english,superscriptaddress,notitlepage]{revtex4-2}
\usepackage[T1]{fontenc}
\usepackage[latin1, utf8]{inputenc}
\usepackage{amsmath}
\usepackage{graphicx}

\makeatletter
\usepackage{hyperref}

\usepackage{bm}
\usepackage{physics}
\usepackage{pifont}
\usepackage{fancyhdr}
\usepackage{color}
\usepackage{textcomp}
\usepackage[toc,page]{appendix}

\hypersetup{hidelinks}
\hypersetup{
colorlinks=true,
linkcolor=blue,
citecolor=blue,
anchorcolor=blue,
urlcolor=blue,
}

\makeatother

\usepackage{babel}
\begin{document}
\title{Cross-phase modulation in the two dimensional spectroscopy}
\author{Mao-Rui Cai}
\affiliation{Graduate School of China Academy of Engineering Physics, No. 10 Xibeiwang
East Road, Haidian District, Beijing 100193, China}
\author{Xue Zhang}
\affiliation{Graduate School of China Academy of Engineering Physics, No. 10 Xibeiwang
East Road, Haidian District, Beijing 100193, China}
\author{Zi-Qian Cheng}
\affiliation{Graduate School of China Academy of Engineering Physics, No. 10 Xibeiwang
East Road, Haidian District, Beijing 100193, China}
\author{Teng-Fei Yan}
\email{yantf@shu.edu.cn}

\affiliation{School of Microelectronics, shanghai university, Shanghai 200444,
China}
\author{Hui Dong}
\email{hdong@gscaep.ac.cn}

\affiliation{Graduate School of China Academy of Engineering Physics, No. 10 Xibeiwang
East Road, Haidian District, Beijing 100193, China}
\begin{abstract}
Developing from the transient absorption (TA) spectroscopy, the two
dimensional (2D) spectroscopy with pump-probe geometry has emerged
as a versatile approach for alleviating the difficulty on implementing
the 2D spectroscopy with other geometries. However, the presence of
cross-phase modulation (XPM) in TA spectroscopy introduces significant
spectral distortions, particularly when the pump and probe pulses
overlap. We demonstrate that this phenomenon is extended to the 2D
spectroscopy with pump-probe geometry and the XPM is induced by the
interference of the two pump pulse. We present the oscillatory behavior
of XPM in the 2D spectrum and its displacement with respect to the
waiting time delay through both experimental measurements and numerical
simulations. Additionally, we explore the influence of probe pulse
chirp on XPM and discover that by compressing the chirp, the impact
of XPM on the desired signal can be reduced.
\end{abstract}
\maketitle

\section{Introduction}

Two dimensional (2D) spectroscopy~\citep{mukamel1995book,cho2009book,Jonas2003,CundiffMukamel}
is a powerful tool with growing attention for studying the couplings
and dynamics of systems in both condensed~\citep{Schlau-Cohen2011,Karaiskaj2010,Reimann2021,Li2021}
and gaseous phases~\citep{Lu2016,Yu2019,Tian2003,Dai2012}. Among
the various 2D spectroscopy beam geometries~\citep{Grumstrup2007,Reimann2021,Li2021},
the pump-probe geometry~\citep{Grumstrup2007,Shim2009,Tekavec2009}
developed from the transient absorption (TA) spectroscopy~\citep{Pollard1989,Yan1990,Berera2009,Ruckebusch2012,Yue2023}
has emerged as a widely adopted approach due to its ease of implementation.
In this technique, the pump beam is typically modulated by a pulse
shaper~\citep{Tekavec2009,Shim2009,Chris2010} which is utilized
to generate two identical pump pulses with interval $\tau$ and to
compensate for the chirp of the pump pulses~\citep{Shim2006}. As
for the probe arm, the chirp of the probe pulse is often compressed
with an extra compressor~\citep{Myers2008} or even left uncompressed,
especially when the super-continuum is applied for probe~\citep{Tekavec2009,Song2021}.
Similar to TA spectroscopy, the chirp of the probe pulse would introduce
distortions~\citep{Megerle2009,Ruckebusch2012,Liebel2015} to the
spectrum in the 2D case~\citep{Schlau-Cohen2011,Tekavec2010,Tekavec2012}.
This occurs because various frequency components of the probe pulse
reach the investigated sample at different times~\citep{Ernsting1999}.
Taking inspiration from the chirp correction scheme used in TA spectroscopy~\citep{Megerle2009,Ruckebusch2012,Liebel2015},
it has been proposed that these distortions in the 2D spectroscopy
could be post-corrected~\citep{Tekavec2010,Tekavec2012} by characterizing
the chirp of the probe pulse. 

Despite the distortion caused by the chirp of the probe pulse, there
is another significant artifact induced by the intense pump pulse
in the TA spectroscopy known as the cross-phase modulation (XPM)~\citep{Ernsting1999,Gardecki2000,Yeremenko2002,Lorenc2002,Ekvall2000}.
XPM universally exists in the TA spectroscopy when the pump and probe
pulses overlap temporally at the sample, particularly in the system
of condensed phase. Moreover, it get stronger as the chirp of the
probe pulse increases~\citep{Ernsting1999}. As an extension of the
TA spectroscopy, the 2D spectroscopy with pump-probe geometry also
gets affected by the XPM~\citep{Park2015} especially when dynamics
within several tens of femtoseconds is investigated. However, a comprehensive
theoretical understanding of the XPM in the 2D spectroscopy with pump-probe
geometry and its response to variations in the chirp of the probe
pulse remain elusive.

In this paper, we present a theoretical derivation of the XPM in the
2D spectroscopy with pump-probe geometry. Our results demonstrate
that XPM manifests itself at the central frequency of the pump pulse
on the excitation axis and at the central frequency of the probe pulse
on the detection axis. Additionally, we observe oscillations and shifts
of the XPM along the detection axis, which are supported by experimental
data using a sapphire window as the sample and by our numerical simulations.
Furthermore, we investigate the effect of probe pulse chirp on the
behavior of XPM through the experiments and simulations. We demonstrated
that by reducing the chirp of the probe pulse, the influence of XPM
on the desired signal is mitigated.

\section{theory}

\subsection{General two dimensional spectroscopy\label{subsec:General-2D}}

\begin{figure}
\includegraphics{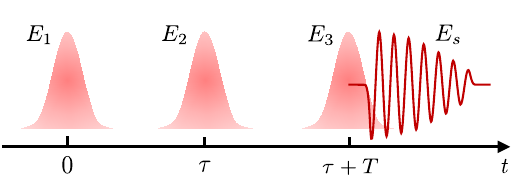}

\caption{Typical pulse sequence of the two dimensional spectroscopy, where
three pulses arrive at the sample at time $t=0$, $t=\tau$, and $t=\tau+T$
sequentially. \label{fig:Pulse-sequence}}
\end{figure}

In typical two dimensional spectroscopy experiments, the third-order
optical response~\citep{mukamel1995book,cho2009book} of a sample
is represented by the induced third-order polarization as
\begin{equation}
P^{(3)}\left(t\right)=\int_{0}^{\infty}dt_{3}\int_{0}^{\infty}dt_{2}\int_{0}^{\infty}dt_{1}R^{(3)}\left(t_{1},t_{2},t_{3}\right)E\left(t-t_{3}\right)E\left(t-t_{3}-t_{2}\right)E\left(t-t_{3}-t_{2}-t_{1}\right).\label{eq:general-response2}
\end{equation}
where $R^{(3)}\left(t_{1},t_{2},t_{3}\right)$ is the third-order
response function of the sample and $E\left(t\right)$ is the external
electric field. As Fig.~\ref{fig:Pulse-sequence} shows, the external
field typically consists of three pulses arriving sequentially at
time $t=0$, $t=\tau$, and $t=\tau+T$~\citep{Chris2010,Yue2015,Schlau-Cohen2011,Zhang2022}
for 2D setup. As a result, both the external electric field and the
induced third-order polarization are functions of $\tau$, $T$ and
$t$. Specifically, we denote the induced third-order polarization
as $P^{(3)}\left(\tau,T,t\right)$, and the electric field as 

\begin{equation}
E\left(\tau,T,t\right)=E_{1}\left(t\right)+E_{2}\left(\tau,t\right)+E_{3}\left(\tau,T,t\right)+\mathrm{c.c.},\label{eq:incident-field}
\end{equation}

\begin{gather}
E_{1}\left(t\right)=A_{1}\left(t\right)\exp\left\{ i\left(\vec{k}_{1}\cdot\vec{r}-\Omega_{1}t+\phi_{1}\right)\right\} ,\\
E_{2}\left(\tau,t\right)=A_{2}\left(t-\tau\right)\exp\left\{ i\left[\vec{k}_{2}\cdot\vec{r}-\Omega_{2}\left(t-\tau\right)+\phi_{2}\right]\right\} ,\\
E_{3}\left(\tau,T,t\right)=A_{3}\left(t-\tau-T\right)\exp\left\{ i\left[\vec{k}_{3}\cdot\vec{r}-\Omega_{3}\left(t-\tau-T\right)+\phi_{3}\right]\right\} .
\end{gather}
where $A_{m}$, $\vec{k}_{m}$, $\Omega_{m}$, and $\phi_{m}$ $(m=1,2,3)$
are respectively the envelopes, wave vectors, central frequencies,
and initial phases of the pulses, and $\vec{r}$ is the spatial location
of the sample molecule. The possible dispersion of the pulses is encoded
into the envelopes. For a transparent sample with length $L$, the
induced polarization yields signal~\citep{mukamel1995book} field
\begin{equation}
E_{s}\left(\tau,T,t\right)=\frac{2\pi i}{n}\frac{\Omega_{s}L}{c}P^{(3)}\left(\tau,T,t\right)=i\eta\Omega_{s}P^{(3)}\left(\tau,T,t\right),
\end{equation}
along several phase matching directions $\vec{k}_{s}=\pm\vec{k}_{\alpha}\pm\vec{k}_{\beta}\pm\vec{k}_{\gamma}$
($\alpha,\beta,\gamma=1,2,3$), where $\eta=2\pi L/cn$, $n$ is the
linear refractive index of the sample, $c$ is the speed of light,
and $\Omega_{s}$ is the central frequency of the signal field. 

In general, the two dimensional (2D) spectrum $E_{s}\left(\omega_{\tau},T,\omega_{t}\right)$
is obtained by Fourier transforming $E_{s}\left(\tau,T,t\right)$
with respect to $\tau$ and $t$. In some cases, the time domain signal
$E_{s}\left(\tau,T,t\right)$ is collected by scanning both $\tau$
and $t$~\citep{Lu2016,Yu2019,Reimann2021,Li2021}. Alternatively,
it is popular to heterodynely detect~\citep{mukamel1995book,Yue2015,Schlau-Cohen2011}
the frequency domain signal $E_{s}\left(\tau,T,\omega_{t}\right)$
via a spectrometer and with a local oscillator $E_{\mathrm{LO}}\left(t\right)$.
In this technique, the differential optical signal $S\left(\tau,T,\omega_{t}\right)$
is obtained as 
\begin{equation}
S\left(\tau,T,\omega_{t}\right)=\ln\frac{I_{s}\left(\tau,T,\omega_{t}\right)}{I_{\mathrm{LO}}\left(\omega_{t}\right)}\simeq\frac{I_{s}\left(\tau,T,\omega_{t}\right)}{I_{\mathrm{LO}}\left(\omega_{t}\right)}-1\simeq\frac{2\mathrm{Re}\left[E_{s}\left(\tau,T,\omega_{t}\right)E_{\mathrm{LO}}^{*}\left(\omega_{t}\right)\right]}{\left|E_{\mathrm{LO}}\left(\omega_{t}\right)\right|^{2}}=2\mathrm{Re}\left[\frac{E_{s}\left(\tau,T,\omega_{t}\right)}{E_{\mathrm{LO}}\left(\omega_{t}\right)}\right],
\end{equation}
where $I_{\mathrm{LO}}\left(\omega_{t}\right)=\left|E_{\mathrm{LO}}\left(\omega_{t}\right)\right|^{2}$
and $I_{s}\left(\tau,T,\omega_{t}\right)=\left|E_{s}\left(\tau,T,\omega_{t}\right)+E_{\mathrm{LO}}\left(\omega_{t}\right)\right|^{2}\simeq2\mathrm{Re}\left[E_{s}\left(\tau,T,\omega_{t}\right)E_{\mathrm{LO}}^{*}\left(\omega_{t}\right)\right]+\left|E_{\mathrm{LO}}\left(\omega_{t}\right)\right|^{2}$,
because the signal field is usually much weaker than the local oscillator,
i.e., $\left|E_{s}\left(\tau,T,\omega_{t}\right)\right|\ll\left|E_{\mathrm{LO}}\left(\omega_{t}\right)\right|$.
The 2D spectrum $S\left(\omega_{\tau},T,\omega_{t}\right)$ is then
obtained by scanning and Fourier transforming with respect to only
$\tau$,
\begin{equation}
S\left(\omega_{\tau},T,\omega_{t}\right)=\mathcal{F}\left[S\left(\tau,T,\omega_{t}\right)\right].
\end{equation}

\subsection{Two dimensional spectroscopy with pump-probe geometry\label{subsec:Pump-probe-2D}}

There are several experimental beam geometries in the 2D spectroscopy
based on the direction arrangement of the three pulses, such as collinear
geometry~\citep{Yu2019,Reimann2021,Li2021} and non-collinear BOXCARS
geometry~\citep{Schlau-Cohen2011,Yue2015}. Besides, there is a partially
collinear geometry known as the pump-probe geometry~\citep{Grumstrup2007,Shim2009,Tekavec2009}
where the first two pulses are collinear $\vec{k}_{1}=\vec{k}_{2}$
and the third pulse serves also as the local oscillator. Such a beam
geometry is identical with the TA (pump-probe) spectroscopy geometry~\citep{Pollard1989,Yan1990,Berera2009,Ruckebusch2012,Yue2023}.
The first two pulses are often denoted as the pump pulses and the
third pulse as the probe pulse. The pump-probe geometry eases the
difficulty of experimentally implementing the 2D spectroscopy compared
with the other geometries, because the heterodyne detection is naturally
satisfied by measuring the spectrum of probe pulse, i.e., $\vec{k}_{s}=\vec{k}_{3}$.
However, signal fields that are spatially separated in the BOXCARS
geometry may get mixed~\citep{Shim2009,Myers2008} in the pump-probe
geometry. Specifically, signal fields $E_{\alpha',-\alpha',3}\left(\tau,T,\omega_{t}\right)$,
$E_{1,-2,3}\left(\tau,T,\omega_{t}\right)$, and $E_{-1,2,3}\left(\tau,T,\omega_{t}\right)$
are all emit along $\vec{k}_{3}$ in the pump-probe geometry. Here,
$\alpha'=\pm1,\pm2,\pm3$ and the subscripts of the signal field $E_{\alpha,\beta,\gamma}$
denote that the signal is induced by incident fields $E_{\alpha}$,
$E_{\beta}$, and $E_{\gamma}$ (or their conjugation if the corresponding
subscripts are negative). Among all these mixed signals, $E_{-1,2,3}\left(\tau,T,\omega_{t}\right)$
and $E_{1,-2,3}\left(\tau,T,\omega_{t}\right)$ are typically required
for the study of single-quantum coherence in the excitation time $\tau$.
These two signals are often denoted as the rephasing and non-rephasing
signal~\citep{mukamel1995book,Schlau-Cohen2011}, respectively.

To sort out only $E_{-1,2,3}\left(\tau,T,\omega_{t}\right)$ and $E_{1,-2,3}\left(\tau,T,\omega_{t}\right)$
in the total signal field, the technique of phase cycling~\citep{Shim2009,Myers2008,Tian2003}
is applied. According to the discussion in Sec.~\ref{subsec:General-2D},
the signal field $E_{s}\left(\tau,T,\omega_{t}\right)$ from the sample
depends on the initial phase $\phi_{\alpha}$ of the incident fields.
However, in the 2D spectroscopy with pump-probe geometry, the phase
of the third pulse $\phi_{3}$ is eliminated because the third pulse
serves as the local oscillator in the heterodyne detection~\citep{Shim2009}.
Thus, we denote the phase-depend signal field as $E_{s}\left(\tau,T,\omega_{t};\phi_{1},\phi_{2}\right)$
and its components 
\begin{equation}
E_{\alpha,\beta,3}\left(\tau,T,\omega_{t};\phi_{1},\phi_{2}\right)=E_{\alpha,\beta,3}\left(\tau,T,\omega_{t}\right)\exp\left\{ i\left[\mathrm{sgn}\left(\alpha\right)\times\phi_{\left|\alpha\right|}+\mathrm{sgn}\left(\beta\right)\times\phi_{\left|\beta\right|}\right]\right\} ,\label{eq:phase-depend-sig}
\end{equation}
where $\alpha,\beta=\pm1,\pm2,\pm3$, $\mathrm{sgn}\left(x\right)=x/\left|x\right|$
is the sign function, and the phase free term $E_{\alpha,\beta,3}\left(\tau,T,\omega_{t}\right)$
denotes the signal field with initial phases $\phi_{1},\phi_{2}=0$
of the first two pulses. By conducting two experiments with phase
arrangements a) $\phi_{1}=\pi$ and $\phi_{2}=0$; b) $\phi_{1}=\pi$
and $\phi_{2}=\pi$, the phase cycled signal is given as 
\begin{equation}
S_{\mathrm{cyc}}\left(\tau,T,\omega_{t}\right)=\ln\frac{I\left(\tau,T,\omega_{t};\pi,0\right)}{I\left(\tau,T,\omega_{t};\pi,\pi\right)}\simeq\frac{I\left(\tau,T,\omega_{t};\pi,0\right)}{I\left(\tau,T,\omega_{t};\pi,\pi\right)}-1\simeq\frac{I\left(\tau,T,\omega_{t};\pi,0\right)-I\left(\tau,T,\omega_{t};\pi,\pi\right)}{I\left(\tau,T,\omega_{t};\pi,\pi\right)},
\end{equation}
with 
\begin{equation}
I\left(\tau,T,\omega_{t};\phi_{1},\phi_{2}\right)=\left|E_{s}\left(\tau,T,\omega_{t};\phi_{1},\phi_{2}\right)+E_{3}(\omega_{t})\right|^{2}\simeq\left|E_{3}(\omega_{t})\right|^{2}+2\mathrm{Re}\left[E_{s}\left(\tau,T,\omega_{t};\phi_{1},\phi_{2}\right)E_{3}^{*}(\omega_{t})\right],
\end{equation}
$I_{3}\left(\omega_{t}\right)=\left|E_{3}(\omega_{t})\right|^{2}$,
and $\left|E_{s}\left(\tau,T,\omega_{t};\phi_{1},\phi_{2}\right)\right|\ll\left|E_{3}(\omega_{t})\right|$.
Then, the phase cycled signal is expressed as
\begin{equation}
S_{\mathrm{cyc}}\left(\tau,T,\omega_{t}\right)=2\mathrm{Re}\left[\frac{E_{s}\left(\tau,T,\omega_{t};\pi,0\right)-E_{s}\left(\tau,T,\omega_{t};\pi,\pi\right)}{E_{3}\left(\omega_{t}\right)}\right],
\end{equation}
which includes only the signals $E_{-1,2,3}$ and $E_{1,-2,3}$, while
the parts $E_{\alpha',-\alpha',3}$ are eliminated because they are
not affected by the initial phases of the pump pulses. The 2D spectrum
$S_{\mathrm{cyc}}\left(\omega_{\tau},T,\omega_{t}\right)$ is obtained
by firstly scanning $\tau$ to collect a series of $S_{\mathrm{cyc}}\left(\tau,T,\omega_{t}\right)$
and then Fourier transforming $S_{\mathrm{cyc}}\left(\tau,T,\omega_{t}\right)$
with respect to $\tau$. The rephasing and non-rephasing parts can
further be separated with additional phase cycling~\citep{Myers2008}.

\subsection{Cross-phase modulation in the two dimensional spectroscopy with pump-probe
geometry}

In most spectroscopy experiments, the frequencies of the incident
electric fields are typically chosen to resonate with (or nearly resonate
with) one of the transition frequencies of the investigated sample,
thereby changing the absorption rate~\citep{mukamel1995book,cho2009book}
(imaginary part of the electric susceptibility) of the sample accordingly.
However, in liquid sample where the solution (i.e., the investigated
sample) is dissolved in a specific solvent, the electric fields also
induce the non-resonant electronic response (ER) of the solvent, altering
the refractive index~\citep{Oudar1983,Park2015} (real part of the
electric susceptibility) of the solvent through Kerr effect. The ER
is approximately instantaneous~\citep{Oudar1983,Hellwarth1975} (with
a time scale of $0.1\sim1\:\mathrm{fs}$) compared with the resonant
response of the solution and the duration of the incident pulses (with
a time scale of $10\sim100\:\mathrm{fs}$). Therefore, the response
function~\citep{Oudar1983,Ernsting1999,Park2015} is approximately
written as 
\begin{equation}
R_{\mathrm{ER}}^{(3)}\left(t_{1},t_{2},t_{3}\right)=\sigma_{e}\delta\left(t_{1}\right)\delta\left(t_{2}\right)\delta\left(t_{3}\right).\label{eq:elec-response}
\end{equation}
Substituting Eq.~(\ref{eq:elec-response}) into Eq.~(\ref{eq:general-response2}),
the induced polarization of the electronic response is given as $P^{\mathrm{ER}}\left(\tau,T,t\right)=\sigma_{e}\left[E\left(\tau,T,t\right)\right]^{3}$,
and the corresponding signal field is 
\begin{equation}
E_{s}^{\mathrm{ER}}\left(\tau,T,t\right)=i\sigma_{e}\eta\Omega_{s}\left[E\left(\tau,T,t\right)\right]^{3}.
\end{equation}

Among all the heterodynely detected signal components in the 2D spectroscopy
with pump-probe geometry, $E_{3,-3,3}^{\mathrm{ER}}\left(\tau,T,t\right)$
and $E_{-3,3,3}^{\mathrm{ER}}\left(\tau,T,t\right)$ are treated as
the self-phase modulation (SPM)~\citep{Alfano1970} which describes
the spectral and temporal modulation of the probe pulse induced by
itself. The other components are cross-phase modulation (XPM)~\citep{Ernsting1999,Gardecki2000,Yeremenko2002,Lorenc2002,Ekvall2000},
describing the spectral and temporal modulation of the probe pulse
induced by the first pump pulse only {[}$E_{1,-1,3}^{\mathrm{ER}}\left(\tau,T,t\right)$
and $E_{-1,1,3}^{\mathrm{ER}}\left(\tau,T,t\right)${]}, the second
pump pulse only {[}$E_{2,-2,3}^{\mathrm{ER}}\left(\tau,T,t\right)$
and $E_{-2,2,3}^{\mathrm{ER}}\left(\tau,T,t\right)${]}, or the interference
of the two pump pulses {[}$E_{-1,2,3}^{\mathrm{ER}}\left(\tau,T,t\right)$
and $E_{1,-2,3}^{\mathrm{ER}}\left(\tau,T,t\right)${]} where
\begin{align}
E_{-1,2,3}^{\mathrm{ER}}\left(\tau,T,t\right) & =i\sigma_{e}\eta\Omega_{s}E_{1}^{*}\left(t\right)E_{2}\left(t-\tau\right)E_{3}\left(t-\tau-T\right),\label{eq:E-XPM-1}\\
E_{1,-2,3}^{\mathrm{ER}}\left(\tau,T,t\right) & =i\sigma_{e}\eta\Omega_{s}E_{1}\left(t\right)E_{2}^{*}\left(t-\tau\right)E_{3}\left(t-\tau-T\right).\label{eq:E-XPM-2}
\end{align}
Via the phase cycling technique~\citep{Shim2009,Myers2008,Tian2003}
described in Sec.~\ref{subsec:Pump-probe-2D}, all the modulations
of the probe pulse induced by a single pulse $E_{\alpha',-\alpha',3}^{\mathrm{ER}}$
are eliminated. Whereas, the modulations from the interference of
the two pump pulses remain,
\begin{equation}
S_{\mathrm{XPM}}\left(\tau,T,\omega_{t}\right)=-4\mathrm{Re}\left[\frac{E_{-1,2,3}^{\mathrm{ER}}\left(\tau,T,\omega_{t}\right)+E_{1,-2,3}^{\mathrm{ER}}\left(\tau,T,\omega_{t}\right)}{E_{3}\left(\omega_{t}\right)}\right],
\end{equation}
where $E_{-1,2,3}^{\mathrm{ER}}\left(\tau,T,\omega_{t}\right)$ and
$E_{1,-2,3}^{\mathrm{ER}}\left(\tau,T,\omega_{t}\right)$ are XPM
signal fields when the initial phases of the two pump pulses $\phi_{1},\phi_{2}=0$.

To give more explicit expressions of the XPM signal $S_{\mathrm{XPM}}\left(\tau,T,\omega_{t}\right)$,
we assume that the two pump pulses are Fourier transform limited Gaussian
pulses without chirp, i.e., 

\begin{gather}
E_{1}\left(t\right)=a_{1}\exp\left\{ -\frac{t^{2}}{2\tau_{1}^{2}}\right\} \exp\left\{ -i\Omega_{1}t\right\} ,\label{eq:E1}\\
E_{2}\left(\tau,t\right)=a_{2}\exp\left\{ -\frac{\left(t-\tau\right)^{2}}{2\tau_{2}^{2}}\right\} \exp\left\{ -i\Omega_{2}\left(t-\tau\right)\right\} ,\label{eq:E2}
\end{gather}
and the probe pulse is a Gaussian pulse with linear chirp~\citep{Ernsting1999,Tekavec2012}
\begin{equation}
E_{3}\left(\tau,T,t\right)=a_{3}\exp\left\{ -\frac{\left(t-\tau-T\right)^{2}}{2\tau_{3}^{2}}\right\} \exp\left\{ -i\left[\Omega_{3}\left(t-\tau-T\right)+\beta_{3}\left(t-\tau-T\right)^{2}\right]\right\} ,\label{eq:E3}
\end{equation}
where $a_{\alpha}$ and $\tau_{\alpha}$ are the amplitude and duration
of pulse $E_{\alpha}$, and $\beta_{3}$ is the chirp rate of the
probe pulse. The spatial phase factors and initial phases are neglected
here. We further simplify our model by assuming $a_{1}=a_{2}$, $\tau_{1}=\tau_{2}$,
and $\Omega_{1}=\Omega_{2}$, which is accessible with the pulse shaper~\citep{Tekavec2009,Shim2009,Chris2010}
on the pump arm. With these incident pulses in Eq.~(\ref{eq:E1}-\ref{eq:E3}),
we obtain the phase cycled XPM signal as 
\begin{equation}
\begin{aligned}S_{\mathrm{XPM}}\left(\tau,T,\omega_{t}\right) & =\frac{8}{A^{1/4}}\sigma_{e}\eta\Omega_{3}\tau_{1}\left|a_{1}\right|^{2}\cos\left(\Omega_{1}\tau\right)\exp\left\{ -\frac{\tau^{2}}{4\tau_{1}^{2}}\right\} \\
 & \times\exp\left\{ \frac{\left(\tau_{1}^{2}+2\tau_{3,0}^{2}\right)B-4\beta_{gdd}\tau_{3,0}^{2}\left(\Omega_{3}-\omega_{t}\right)t_{d}\left(\omega_{t}\right)}{A}\right\} \\
 & \times\sin\left\{ \frac{2\beta_{gdd}B+2\tau_{3,0}^{2}\left(\Omega_{3}-\omega_{t}\right)\left(\tau_{1}^{2}+2\tau_{3,0}^{2}\right)t_{d}\left(\omega_{t}\right)}{A}+\phi\right\} ,
\end{aligned}
\label{eq:sig-XPM-1}
\end{equation}
where $A=\left(\tau_{1}^{2}+2\tau_{3,0}^{2}\right)^{2}+4\beta_{gdd}^{2}$,
$B=\left[t_{d}\left(\omega_{t}\right)\right]^{2}+\tau_{3,0}^{4}\left(\Omega_{3}-\omega_{t}\right)^{2}$,
$t_{d}\left(\omega_{t}\right)=\tau/2+T-\beta_{gdd}\left(\Omega_{3}-\omega_{t}\right)$,
and $\phi=\arctan\left[2\beta_{gdd}/\left(\tau_{1}^{2}+2\tau_{3,0}^{2}\right)\right]$.
The parameters $\beta_{gdd}$ and $\tau_{3,0}$ are the group delay
velocity (GDD) and Fourier limited duration of the probe pulse, respectively.
These two parameters are connected  to $\tau_{3}$ and $\beta_{3}$
with relation $\tau_{3}^{2}=\left(\tau_{3,0}^{4}+\beta_{gdd}^{2}\right)/\tau_{3,0}^{2}$
and $\beta_{3}=0.5\beta_{gdd}/\left(\tau_{3,0}^{4}+\beta_{gdd}^{2}\right)$~\citep{Tekavec2012}. 

The cosine term $\cos\left(\Omega_{1}\tau\right)$ and the Gaussian
term with variable $\left(\Omega_{3}-\omega_{t}\right)$ in Eq.~(\ref{eq:sig-XPM-1})
indicate that the XPM signal on the 2D spectrum covers around $\left(\omega_{\tau}=\Omega_{1},\omega_{t}=\Omega_{3}\right)$,
i.e., typically the resonant frequencies of the sample solution. We
remark that the XPM signal are oscillating along $\omega_{t}$ govern
by the sine term in Eq.~(\ref{eq:sig-XPM-1}), and the oscillation
will move along $\omega_{t}$ as $T$ changes. We will show such features
with experiments and simulations in the following sections.

\section{Experiment and simulation}

\subsection{Experimental Setup}

\begin{figure}[tbph]
\includegraphics{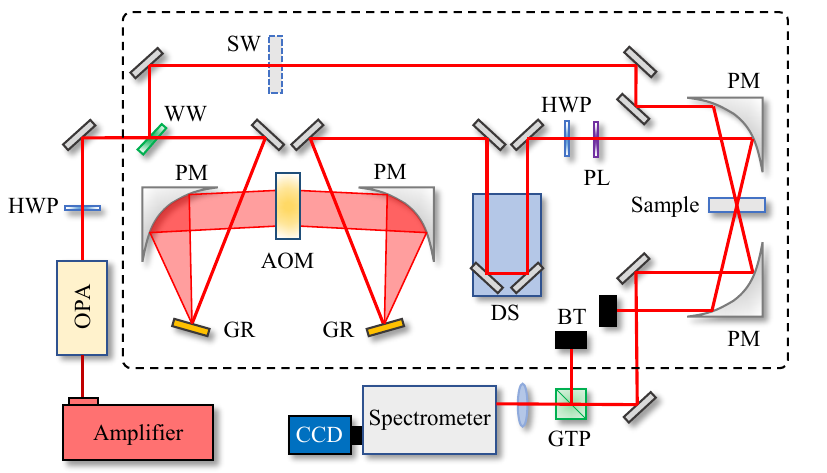}

\caption{Schematic experimental setup of the 2D spectroscopy with pump-probe
geometry in our experiment. Amplifier: Ti:Sapphire amplifier laser
system; OPA: optical parametric amplifier; HWP: half-wave plate; WW:
wedged window; PM: parabolic mirror; GR: optical grating; AOM: acousto-optic
modulator; DS: delay stage; PL: polarizer; SW: sapphire window (3
mm) used to altering the GDD of the probe pulse; BT: beam trap; GTP:
Glan-Taylor prism; Spectrometer: spectrometer; CCD: charge coupled
device camera. The system enclosed by a dash box is an integrated
2D spectroscopy system. In our 2D spectroscopy experiment for detecting
the XPM, a sapphire window (3 mm) serves as the sample.\label{fig:EXP-setup}}

\end{figure}

Our experimental setup is presented in Fig.~\ref{fig:EXP-setup}.
Laser pulses (800 nm, 7.15 W, 30.8 fs, at a 1 kHz repetition rate,
and horizontally polarized with respect to the table surface) delivered
by the Ti:Sapphire amplifier laser system are lead into an optical
parametric amplifier (OPA) to produce pulses with desired central
wavelength at 680 nm. The laser beam is then lead into a commercial
integrated 2D spectroscopy system. Specifically, the laser beam after
the OPA is divided by a wedged window into two beams, i.e., the pump
and probe beams, and the power of the probe beam is controlled by
a half-wave plate before the wedged window. To introduce time delays
between the pump and probe pulses, the pump beam is directed into
a pulse shaper and a delay stage before hitting the sample. The pulse
shaper~\citep{Tekavec2009,Shim2009,Chris2010} consists of an acousto-optic
modulator (AOM), two parabolic mirrors, and two gratings with 4-f
geometry. The AOM, synchronized with our laser system, modulates the
first-order diffraction of the pump pulses with customized mask through
photoelastic effect. Typically, the mask is designed such that one
pump pulse is divided into two identical transform limited pump pulses
with interval $\tau$. The time delay $T$ between the probe and the
second pump pulse is controlled by the delay stage. The pump and probe
beams are aligned in parallel and focused onto the sample using a
parabolic mirror. After the sample, the pump beam is blocked by a
beam trap, and the probe beam, passing through a Glan-Taylor prism
and a lens ($f=300\:\mathrm{mm}$), is lead into a spectrometer with
the spectrum detected by a charge coupled device (CCD) camera. The
Glan-Taylor prism is used to guarantee that only horizontally polarized
component of the probe pulse is measured. Furthermore, a half-wave
plate and a polarizer in the pump arm before the sample are utilized
to ensure that only horizontally polarized component pump pulses interact
with the sample. 

In our 2D spectroscopy experiment for detecting the XPM, a sapphire
window (3 mm) serves as the sample. Additionally, to study the influence
of the GDD of the probe pulse to the XPM on the 2D spectrum, two experiments
are conducted. In our first experiment, the probe beam passes freely
before the sample, while in the second experiment, an additional sapphire
window (SW, 3 mm thick) is placed on probe arm to alter the $\beta_{gdd}$
of the probe pulse. 

We remark that the phases of the pump pulses are also controlled by
the AOM, making phase cycling available in this system. Moreover,
instead of the two-frame cycling $(\phi_{1}=\pi,\phi_{2}=0)$ and
$(\phi_{1}=\pi,\phi_{2}=\pi)$ introduced in Sec.~\ref{subsec:Pump-probe-2D},
we actually utilize a four-frame cycling~\citep{Shim2009} $(\phi_{1}=\pi,\phi_{2}=0)$,
$(\phi_{1}=\pi,\phi_{2}=\pi)$, $(\phi_{1}=0,\phi_{2}=\pi)$, and
$(\phi_{1}=0,\phi_{2}=0)$, which further eliminates the scatter from
the pump pulses.

\subsection{Pulse parameters\label{subsec:Pulse-parameters}}

\begin{figure}[tbph]
\includegraphics{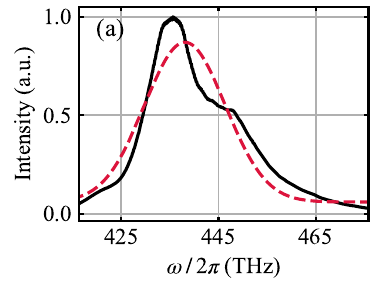}\hspace{1cm}\includegraphics{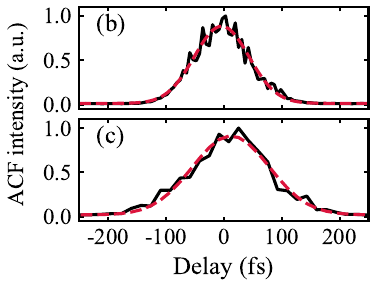}

\caption{(a) The spectrum and (b, c) the auto-correlation functions (ACFs)
of the probe pulses. The black solid lines are experimental results
and the red dash lines are Gaussian fittings. The Gaussian fitting
in (a) suggests that the probe spectrum centered at $\Omega_{3}/2\pi=438.09\:\mathrm{THz}$
with the bandwidth $\sigma_{3,0}/2\pi=11.74\:\mathrm{THz}$. Such
a bandwidth corresponds to a Fourier limited duration $\tau_{3,0}=1/\sigma_{3,0}=13.56\:\mathrm{fs}$
of the probe pulse. The ACFs in (b) and (c) relate to the probe pulses
without and with the additional SW, respectively. Via the Gaussian
fitting, the duration of the ACFs are $\tau_{\mathrm{ACF}}^{e_{1}}=47.25\:\mathrm{fs}$
for (b) and $\tau_{\mathrm{ACF}}^{e_{2}}=66.50\:\mathrm{fs}$ for
(c) corresponding to the duration $\tau_{3}^{e_{1}}=\tau_{\mathrm{ACF}}^{e_{1}}=47.25\:\mathrm{fs}$
and $\tau_{3}^{e_{2}}=\tau_{\mathrm{ACF}}^{e_{2}}=66.50\:\mathrm{fs}$
for the probe pulses. \label{fig:probe-param}}
\end{figure}

We presented in this section the parameters of the pulses applied
in our experiments. The spectrum of the probe pulse is shown in Fig.~\ref{fig:probe-param}(a),
with the central frequency $\Omega_{3}/2\pi=438.09\:\mathrm{THz}$
and the bandwidth $\sigma_{3,0}/2\pi=11.74\:\mathrm{THz}$ evaluated
by the Gaussian fitting of the spectrum. The bandwidth $\sigma_{3,0}$
corresponds to a Fourier limited duration of the probe pulse, $\tau_{3,0}=1/\sigma_{3,0}=13.56\:\mathrm{fs}$.
The auto-correlation functions (ACF) of the probe pulses $E_{3}^{e_{1}}$
and $E_{3}^{e_{2}}$ are given in Fig.~\ref{fig:probe-param}(b)
and (c). Here, the superscripts $e_{1}$ and $e_{2}$ are used to
denote experiments without and with the additional SW on the probe
arm, respectively. The duration of the ACFs are $\tau_{\mathrm{ACF}}^{e_{1}}=47.25\:\mathrm{fs}$
and $\tau_{\mathrm{ACF}}^{e_{2}}=66.50\:\mathrm{fs}$, and the duration
of the electric fields are directly given by the ACFs as $\tau_{3}^{e_{1}}=\tau_{\mathrm{ACF}}^{e_{1}}=47.25\:\mathrm{fs}$
and $\tau_{3}^{e_{2}}=\tau_{\mathrm{ACF}}^{e_{2}}=66.50\:\mathrm{fs}$.
Such duration of the electric fields corresponds to the typical positive
GDDs $\beta_{gdd}^{e_{1}}=613.76\:\mathrm{fs}^{2}$ and $\beta_{gdd}^{e_{2}}=882.79\:\mathrm{fs}^{2}$
according to the relation $\beta_{gdd}^{2}=\tau_{3,0}^{2}\left(\tau_{3}^{2}-\tau_{3,0}^{2}\right)$.
The spectrum is collected by the spectrometer and CCD camera, and
the ACFs are measured by an auto-correlator.

The parameters of the pump pulses are not measured with the spectrometer,
instead we evaluate the parameters of the pump pulses by simulating
the 2D correlation spectra of the XPM (referred to as 2DCS-XPM) and
Gaussian fitting their projection traces on the axis of $\omega_{\tau}$.
Firstly, the central frequencies of the pump pulses are determined
by projecting the experimental results of the 2DCS-XPM onto the axis
of $\omega_{\tau}$. Then the simulations are conducted by scanning
the duration of the pump pulses to obtain the best duration, with
which the simulated projection trace fits best with the experiment
(See Supplementary for details). The central frequencies and best
duration of the pump pulses are $\Omega_{1}^{e_{1}}/2\pi=449.59\:\mathrm{THz}$
and $\tau_{1}^{e_{1}}=48.85\:\mathrm{fs}$ for experiment without
the SW and $\Omega_{1}^{e_{2}}/2\pi=449.19\:\mathrm{THz}$ and $\tau_{1}^{e_{2}}=50.48\:\mathrm{fs}$
for experiment with the SW. These parameters are utilized in Sec.~\ref{subsec:Simulation}
for numerical simulation.

\subsection{XPM in 2D spectroscopy experiments}

\begin{figure}[tbph]
\includegraphics{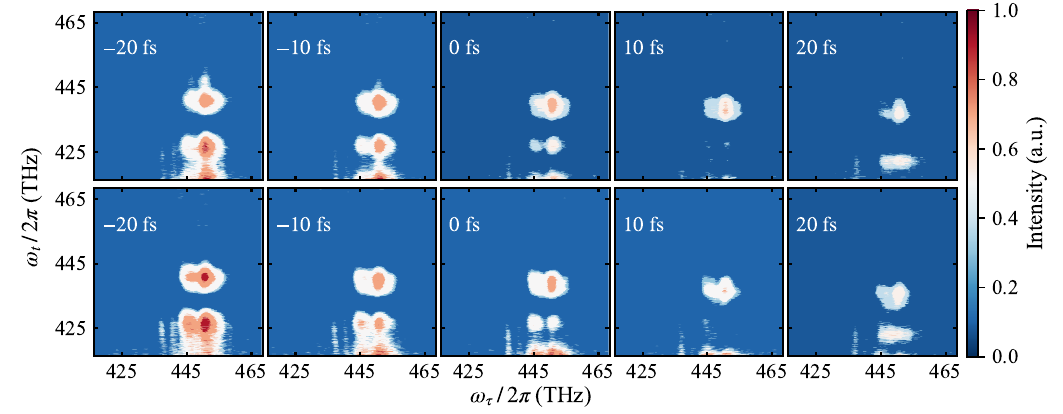}

\caption{2D correlation spectra of the XPM by using a sapphire window as the
sample with the corresponding $T$ denoted on. The spectra in the
first row are measured when the SW is absent on the probe arm. The
spectra in the second row are measured with the the SW placed. Those
2D spectra are obtained by Fourier transform the time domain signal
$S_{\mathrm{XPM}}\left(\tau,T,\omega_{t}\right)$ with respect to
$\tau$ and then taking the absolute value of the transform results.
To stress the signal from the background and from the noise, 65\%
contours are shown here.\label{fig:2D-correlation-spectra}}
\end{figure}

We have measured the 2DCS-XPM using a sapphire window as the sample.
The time delay $T$ is scanned from -100 fs to 30 fs with a step size
of 10 fs. The results from $T=-20\;\mathrm{fs}$ to $T=20\;\mathrm{fs}$
with 65\% contours are shown in Fig.~\ref{fig:2D-correlation-spectra},
where the spectra in the first row are measured without the SW and
those in the second row are measured with the SW on the probe arm.

In Fig.~\ref{fig:2D-correlation-spectra}, up to three peaks along
$\omega_{t}$ are observable. Those peaks correspond to the oscillation
of the $S_{\mathrm{XPM}}\left(\tau,T,\omega_{t}\right)$ introduced
by the sine term in Eq.~(\ref{eq:sig-XPM-1}). The additional peaks,
which are also predicted by the sine term, are not visible here constrained
by the signal-to-noise ratio. We remark that the intensities of the
2DCS-XPM presented in Fig.~\ref{fig:2D-correlation-spectra} are
stronger when $T$ is negative than the positive ones. This discrepancy
arises due to the greater likelihood of the probe pulse overlapping
with the interference of the two pump pulses under negative $T$ conditions. 

\begin{figure}[tbph]
\includegraphics{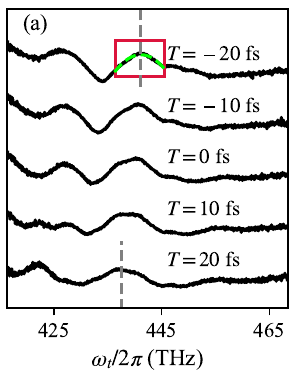} \includegraphics{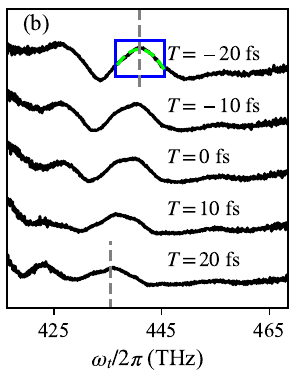}\quad{}\includegraphics{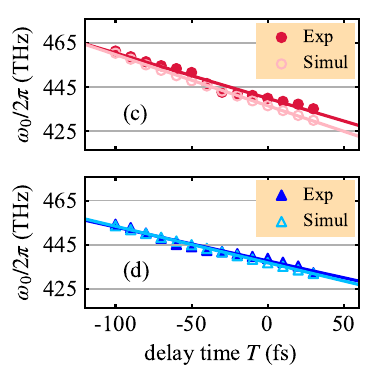}

\caption{(a, b) Projection traces of the 2DCS-XPM onto the axis of $\omega_{t}$
and (c, d) the dynamic shifts of the XPM along the axis of $\omega_{t}$
as $T$ changes. The traces in (a) and the shifts in (c) correspond
to the 2DCS-XPM without the SW and those in (b) and (d) correspond
to the 2DCS-XPM with the SW on the probe arm. The displacement of
the traces in (a) and (b) are indicated by the gray dash lines, representing
the central frequencies of one specific peak, e.g., the one enclosed
by red square in (a) or blue square in (b) when $T=-20\;\mathrm{fs}$.
The central frequencies are obtained by fitting the peaks using the
quadratic polynomial (green dash lines). The shifts of the central
frequencies as functions of $T$ are quantified in (c) by the red
dots and in (d) by the blue solid triangles. Their corresponding linear
fitting lines have slopes of $-0.205\:\mathrm{THz}/\mathrm{fs}$ and
$-0.153\:\mathrm{THz}/\mathrm{fs}$, respectively. Additionally, the
simulation results are depicted in (c) by the pink circles and in
(d) by the light-blue hollow triangles, with the slopes of their linear
fitting lines being $-0.231\:\mathrm{THz}/\mathrm{fs}$ and $-0.165\:\mathrm{THz}/\mathrm{fs}$,
respectively. \label{fig:EXP-Projection}}
\end{figure}

Another feature of the 2DCS-XPM is the dynamic shift along $\omega_{t}$
from higher frequency (shorter wavelength) to smaller frequency (longer
wavelength) regions as the delay time $T$ increases. We emphasize
such a tendency by projecting the 2DCS-XPM onto the axis of $\omega_{t}$,
and the instances from $T=-20\;\mathrm{fs}$ to $T=20\;\mathrm{fs}$
are presented in Fig.~\ref{fig:EXP-Projection}(a) for experiment
without the SW and Fig.~\ref{fig:EXP-Projection}(b) for experiment
with the SW. After the projection, one particular peak near $\Omega_{3}$
of the projection traces, e.g., the one enclosed by red square in
Fig.~\ref{fig:EXP-Projection}(a) or blue square in Fig.~\ref{fig:EXP-Projection}(b),
is fitted with the quadratic polynomial to obtain a central frequency
of the peak. Fig.~\ref{fig:EXP-Projection}(c) and (d) illustrates
the shifts of the central frequencies as $T$ changes from -100 fs
to 30 fs with the red dots (for experiment without the SW) and the
blue triangles (for experiment with the SW), respectively. The shifts
are nearly linear about $T$ in both experiments and the slopes of
the linear fitting lines are $-0.205\:\mathrm{THz}/\mathrm{fs}$ for
experiment without the SW and $-0.153\:\mathrm{THz}/\mathrm{fs}$
for experiment with the SW. The slope characterizes the speed of the
peak displacement, and thus the result shows that the XPM moves faster
when the probe pulse is less chirped.

\subsection{Simulation\label{subsec:Simulation}}

\begin{figure}[tbph]
\includegraphics{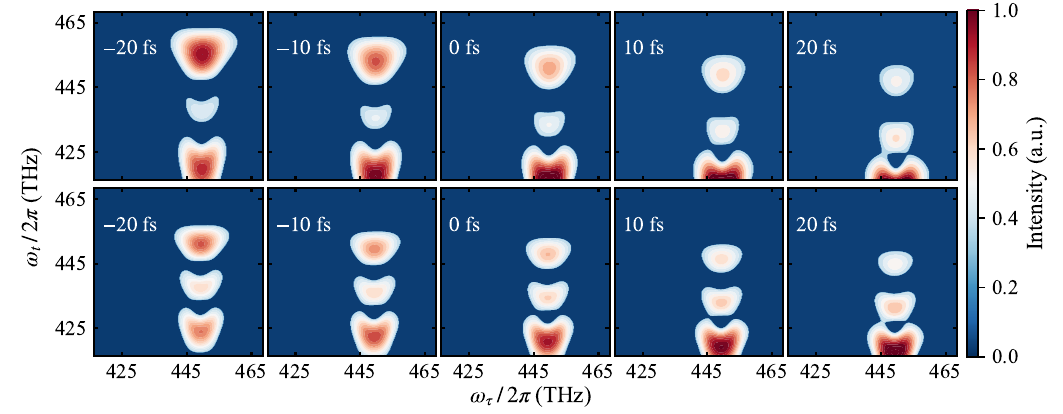}

\caption{Simulated 2DCS-XPM with the corresponding $T$ denoted on. The spectra
in the first row are simulated for the experiment without the SW,
and the parameters are $\Omega_{3}/2\pi=438.09\:\mathrm{THz}$, $\beta_{gdd}^{e_{1}}=613.76\:\mathrm{fs}^{2}$
for the probe pulse, and $\Omega_{\tau}^{e_{1}}/2\pi=449.59\:\mathrm{THz}$,
$\tau_{1}^{e_{1}}=48.85\:\mathrm{fs}$ for the pump pulses. The spectra
in the second row are simulated for the experiment with the SW, and
the parameters are $\Omega_{3}/2\pi=438.09\:\mathrm{THz}$, $\beta_{gdd}^{e_{2}}=882.79\:\mathrm{fs}^{2}$
for the probe pulse, and $\Omega_{1}^{e_{2}}/2\pi=449.19\:\mathrm{THz}$,
$\tau_{1}^{e_{2}}=50.48\:\mathrm{fs}$ for the pump pulses. 65\% contours
are shown here.\label{fig:Simulated-2DCS-XPM}}
\end{figure}

To verify the mechanism of the XPM on the 2D spectroscopy, we simulate
the 2DCS-XPM using Eq.~(\ref{eq:sig-XPM-1}) and with the pulse parameters
in Sec.~\ref{subsec:Pulse-parameters}. The simulation results from
$T=-20\;\mathrm{fs}$ to $T=20\;\mathrm{fs}$ with 65\% contours are
taken as the instances in Fig.~\ref{fig:Simulated-2DCS-XPM}, where
three distinct peaks are observed as in the experimental results in
Fig.~\ref{fig:2D-correlation-spectra}. Also, as in Fig.~\ref{fig:2D-correlation-spectra},
simulations in the first row in Fig.~\ref{fig:Simulated-2DCS-XPM}
correspond to experiment without the SW, and those in the second row
correspond to experiment with the SW. 

We also illustrate the dynamic shifts of the XPM on the simulated
2D spectra by projecting them onto the axis of $\omega_{t}$ and then
fitting the peak near $\Omega_{3}$ to obtain a central frequency.
The shifts of the central frequencies in our simulations are depicted
in Fig.~\ref{fig:EXP-Projection}(c) and (d), alongside the corresponding
experimental results. In Fig.~\ref{fig:EXP-Projection}(c) and (d),
the displacements of the simulations (represented by the hollow marks)
meet well with the experimental results (represented by the solid
marks) for both experiments without and with the SW. The slope of
the linear fittings of the simulations are $-0.231\:\mathrm{THz}/\mathrm{fs}$
and $-0.165\:\mathrm{THz}/\mathrm{fs}$, close to the experimental
results $-0.205\:\mathrm{THz}/\mathrm{fs}$ and $-0.153\:\mathrm{THz}/\mathrm{fs}$.
Both experiments and simulations prove that as the chirp of the probe
pulse gets larger, the XPM moves slower with respect to delay time
$T$ and will cover the desired signal for a wider range of $T$. 

\begin{figure}
\includegraphics{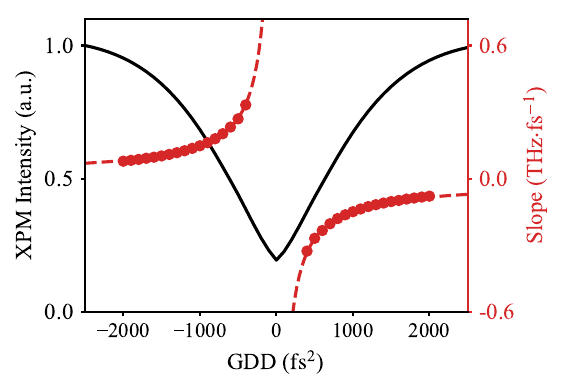}

\caption{Simulated intensities and moving speed of the XPM on the 2D spectra
as functions of the GDD of the probe pulse. In the simulation, the
central frequency and the Fourier limited duration of the probe pulse
are $\Omega_{3}/2\pi=438.09\:\mathrm{THz}$ and $\tau_{3,0}=13.56\:\mathrm{fs}$,
and the central frequency and the Fourier limited duration of the
pump pulse are $\Omega_{1}/2\pi=449.46\:\mathrm{THz}$ and $\tau_{1}=50\:\mathrm{fs}$.
The maximum intensity of the XPM within the region $\left\{ \left(\omega_{\tau},\omega_{t}\right)|444.46\:\mathrm{THz}\protect\leq\omega_{\tau}/2\pi\protect\leq454.46\:\mathrm{THz},433.09\:\mathrm{THz}\protect\leq\omega_{t}/2\pi\protect\leq443.09\:\mathrm{THz}\right\} $
at $T=0\:\mathrm{fs}$ is illustrated by the black solid line. The
moving speed, i.e., the slope, of the XPM with respect to $T$ is
represented by the red solid marks, which are then fitted with reciprocal
function by the red dash lines.\label{fig:intens-slope}}
\end{figure}

The speed of the XPM displacement as a function of the GDD of the
probe pulse $\beta_{gdd}$ is presented in Fig.~\ref{fig:intens-slope},
where the slopes for both positive and negative $\beta_{gdd}$ are
simulated (red solid marks) and then fitted (red dash lines) with
reciprocal function. Unlike in the positive $\beta_{gdd}$ region
where the slope is negative and the XPM moves from higher frequency
(shorter wavelength) to smaller frequency (longer wavelength), in
the negative $\beta_{gdd}$ region, slope is positive and the XPM
moves from smaller frequency (longer wavelength) to larger frequency
(shorter wavelength). Whereas, the reciprocal relation between the
slope and the $\beta_{gdd}$ retains. Such a reciprocal relation moves
the XPM away from the frequency region of interest much faster with
respect to $T$ as $\beta_{gdd}$ decreases. Hence, suppressing the
chirp of the probe pulse also suppresses the range of $T$ within
which the desired signal is affected by the XPM. Moreover, the intensity
of the XPM at $T=0\:\mathrm{fs}$ is significantly reduced by suppressing
$\beta_{gdd}$, as depicted by the black solid line in Fig.~\ref{fig:intens-slope}.
With these two features, one is able to minimize the influence of
the XPM on the 2D spectroscopy by suppressing the chirp of the probe
pulse.

\section{Conclusion}

We have successfully demonstrated the presence of XPM induced by the
interference of the two pump pulses in 2D spectroscopy with pump-probe
geometry. This XPM phenomenon predominantly occurs at $\left(\omega_{\tau}=\Omega_{1},\omega_{t}=\Omega_{3}\right)$,
which typically corresponds to the resonant frequencies of the target
sample. Consequently, the XPM effect can significantly overlap with
the desired signal on the 2D spectra, especially when studying liquid
samples due to the substantial XPM introduced by the solvent.

We have present theoretically and experimentally the pattern and the
displacements of the XPM on the 2D spectrum. We observe that the XPM
oscillates along $\omega_{t}$, while the observable oscillations
(or peaks) are limited by the signal-to-noise ratio. In our experiments,
three peaks are observed. As for the displacements of the XPM, we
show that for positive chirped probe pulse, the XPM on the 2D spectra
moves from higher frequency (shorter wavelength) to smaller frequency
(longer wavelength) regions as $T$ increases. The direction of such
a displacement is reversed for negative chirped probe pulse. Additionally,
we determine that the speed of the displacement with respect to $T$
is inversely related to the GDD of the probe pulse $\beta_{gdd}$.
When the $\beta_{gdd}$ gets suppressed, the XPM moves faster and
thus will disturb the measurement of the desired signal for a narrow
range of $T$. Moreover, suppressing the $\beta_{gdd}$ of the probe
pulse reduces the intensity of the XPM.

\addcontentsline{toc}{section}{\refname}\bibliography{TwoDES_XPM_bib}

\begin{thebibliography}{39}%
\makeatletter
\providecommand \@ifxundefined [1]{%
 \@ifx{#1\undefined}
}%
\providecommand \@ifnum [1]{%
 \ifnum #1\expandafter \@firstoftwo
 \else \expandafter \@secondoftwo
 \fi
}%
\providecommand \@ifx [1]{%
 \ifx #1\expandafter \@firstoftwo
 \else \expandafter \@secondoftwo
 \fi
}%
\providecommand \natexlab [1]{#1}%
\providecommand \enquote  [1]{``#1''}%
\providecommand \bibnamefont  [1]{#1}%
\providecommand \bibfnamefont [1]{#1}%
\providecommand \citenamefont [1]{#1}%
\providecommand \href@noop [0]{\@secondoftwo}%
\providecommand \href [0]{\begingroup \@sanitize@url \@href}%
\providecommand \@href[1]{\@@startlink{#1}\@@href}%
\providecommand \@@href[1]{\endgroup#1\@@endlink}%
\providecommand \@sanitize@url [0]{\catcode `\\12\catcode `\$12\catcode `\&12\catcode `\#12\catcode `\^12\catcode `\_12\catcode `\%12\relax}%
\providecommand \@@startlink[1]{}%
\providecommand \@@endlink[0]{}%
\providecommand \url  [0]{\begingroup\@sanitize@url \@url }%
\providecommand \@url [1]{\endgroup\@href {#1}{\urlprefix }}%
\providecommand \urlprefix  [0]{URL }%
\providecommand \Eprint [0]{\href }%
\providecommand \doibase [0]{https://doi.org/}%
\providecommand \selectlanguage [0]{\@gobble}%
\providecommand \bibinfo  [0]{\@secondoftwo}%
\providecommand \bibfield  [0]{\@secondoftwo}%
\providecommand \translation [1]{[#1]}%
\providecommand \BibitemOpen [0]{}%
\providecommand \bibitemStop [0]{}%
\providecommand \bibitemNoStop [0]{.\EOS\space}%
\providecommand \EOS [0]{\spacefactor3000\relax}%
\providecommand \BibitemShut  [1]{\csname bibitem#1\endcsname}%
\let\auto@bib@innerbib\@empty
\bibitem [{\citenamefont {Mukamel}(1995)}]{mukamel1995book}%
  \BibitemOpen
  \bibfield  {author} {\bibinfo {author} {\bibfnamefont {S.}~\bibnamefont {Mukamel}},\ }\href@noop {} {\emph {\bibinfo {title} {Principles of Nonlinear Optical Spectroscopy}}}\ (\bibinfo  {publisher} {Oxford University Press},\ \bibinfo {year} {1995})\BibitemShut {NoStop}%
\bibitem [{\citenamefont {Cho}(2009)}]{cho2009book}%
  \BibitemOpen
  \bibfield  {author} {\bibinfo {author} {\bibfnamefont {M.}~\bibnamefont {Cho}},\ }\href@noop {} {\emph {\bibinfo {title} {Two-Dimensional Optical Spectroscopy}}}\ (\bibinfo  {publisher} {CRC Press},\ \bibinfo {year} {2009})\BibitemShut {NoStop}%
\bibitem [{\citenamefont {Jonas}(2003)}]{Jonas2003}%
  \BibitemOpen
  \bibfield  {author} {\bibinfo {author} {\bibfnamefont {D.~M.}\ \bibnamefont {Jonas}},\ }\bibfield  {title} {\bibinfo {title} {Two-dimensional femtosecond spectroscopy},\ }\href {https://doi.org/10.1146/annurev.physchem.54.011002.103907} {\bibfield  {journal} {\bibinfo  {journal} {Annual Review of Physical Chemistry}\ }\textbf {\bibinfo {volume} {54}},\ \bibinfo {pages} {425} (\bibinfo {year} {2003})},\ \bibinfo {note} {pMID: 12626736}\BibitemShut {NoStop}%
\bibitem [{\citenamefont {Cundiff}\ and\ \citenamefont {Mukamel}(2013)}]{CundiffMukamel}%
  \BibitemOpen
  \bibfield  {author} {\bibinfo {author} {\bibfnamefont {S.~T.}\ \bibnamefont {Cundiff}}\ and\ \bibinfo {author} {\bibfnamefont {S.}~\bibnamefont {Mukamel}},\ }\bibfield  {title} {\bibinfo {title} {Optical multidimensional coherent spectroscopy},\ }\href {https://doi.org/10.1063/PT.3.2047} {\bibfield  {journal} {\bibinfo  {journal} {Physics Today}\ }\textbf {\bibinfo {volume} {66}},\ \bibinfo {pages} {44} (\bibinfo {year} {2013})}\BibitemShut {NoStop}%
\bibitem [{\citenamefont {Schlau-Cohen}\ \emph {et~al.}(2011)\citenamefont {Schlau-Cohen}, \citenamefont {Ishizaki},\ and\ \citenamefont {Fleming}}]{Schlau-Cohen2011}%
  \BibitemOpen
  \bibfield  {author} {\bibinfo {author} {\bibfnamefont {G.~S.}\ \bibnamefont {Schlau-Cohen}}, \bibinfo {author} {\bibfnamefont {A.}~\bibnamefont {Ishizaki}},\ and\ \bibinfo {author} {\bibfnamefont {G.~R.}\ \bibnamefont {Fleming}},\ }\bibfield  {title} {\bibinfo {title} {Two-dimensional electronic spectroscopy and photosynthesis: Fundamentals and applications to photosynthetic light-harvesting},\ }\href {https://doi.org/https://doi.org/10.1016/j.chemphys.2011.04.025} {\bibfield  {journal} {\bibinfo  {journal} {Chemical Physics}\ }\textbf {\bibinfo {volume} {386}},\ \bibinfo {pages} {1} (\bibinfo {year} {2011})}\BibitemShut {NoStop}%
\bibitem [{\citenamefont {Karaiskaj}\ \emph {et~al.}(2010)\citenamefont {Karaiskaj}, \citenamefont {Bristow}, \citenamefont {Yang}, \citenamefont {Dai}, \citenamefont {Mirin}, \citenamefont {Mukamel},\ and\ \citenamefont {Cundiff}}]{Karaiskaj2010}%
  \BibitemOpen
  \bibfield  {author} {\bibinfo {author} {\bibfnamefont {D.}~\bibnamefont {Karaiskaj}}, \bibinfo {author} {\bibfnamefont {A.~D.}\ \bibnamefont {Bristow}}, \bibinfo {author} {\bibfnamefont {L.}~\bibnamefont {Yang}}, \bibinfo {author} {\bibfnamefont {X.}~\bibnamefont {Dai}}, \bibinfo {author} {\bibfnamefont {R.~P.}\ \bibnamefont {Mirin}}, \bibinfo {author} {\bibfnamefont {S.}~\bibnamefont {Mukamel}},\ and\ \bibinfo {author} {\bibfnamefont {S.~T.}\ \bibnamefont {Cundiff}},\ }\bibfield  {title} {\bibinfo {title} {Two-quantum many-body coherences in two-dimensional fourier-transform spectra of exciton resonances in semiconductor quantum wells},\ }\href {https://doi.org/10.1103/PhysRevLett.104.117401} {\bibfield  {journal} {\bibinfo  {journal} {Phys. Rev. Lett.}\ }\textbf {\bibinfo {volume} {104}},\ \bibinfo {pages} {117401} (\bibinfo {year} {2010})}\BibitemShut {NoStop}%
\bibitem [{\citenamefont {Reimann}\ \emph {et~al.}(2021)\citenamefont {Reimann}, \citenamefont {Woerner},\ and\ \citenamefont {Elsaesser}}]{Reimann2021}%
  \BibitemOpen
  \bibfield  {author} {\bibinfo {author} {\bibfnamefont {K.}~\bibnamefont {Reimann}}, \bibinfo {author} {\bibfnamefont {M.}~\bibnamefont {Woerner}},\ and\ \bibinfo {author} {\bibfnamefont {T.}~\bibnamefont {Elsaesser}},\ }\bibfield  {title} {\bibinfo {title} {Two-dimensional terahertz spectroscopy of condensed-phase molecular systems},\ }\bibfield  {journal} {\bibinfo  {journal} {The Journal of Chemical Physics}\ }\textbf {\bibinfo {volume} {154}},\ \href {https://doi.org/10.1063/5.0046664} {10.1063/5.0046664} (\bibinfo {year} {2021}),\ \bibinfo {note} {120901}\BibitemShut {NoStop}%
\bibitem [{\citenamefont {Li}\ \emph {et~al.}(2021)\citenamefont {Li}, \citenamefont {Trovatello}, \citenamefont {Dal~Conte}, \citenamefont {Nuß}, \citenamefont {Soavi}, \citenamefont {Wang}, \citenamefont {Ferrari}, \citenamefont {Cerullo},\ and\ \citenamefont {Brixner}}]{Li2021}%
  \BibitemOpen
  \bibfield  {author} {\bibinfo {author} {\bibfnamefont {D.}~\bibnamefont {Li}}, \bibinfo {author} {\bibfnamefont {C.}~\bibnamefont {Trovatello}}, \bibinfo {author} {\bibfnamefont {S.}~\bibnamefont {Dal~Conte}}, \bibinfo {author} {\bibfnamefont {M.}~\bibnamefont {Nuß}}, \bibinfo {author} {\bibfnamefont {G.}~\bibnamefont {Soavi}}, \bibinfo {author} {\bibfnamefont {G.}~\bibnamefont {Wang}}, \bibinfo {author} {\bibfnamefont {A.~C.}\ \bibnamefont {Ferrari}}, \bibinfo {author} {\bibfnamefont {G.}~\bibnamefont {Cerullo}},\ and\ \bibinfo {author} {\bibfnamefont {T.}~\bibnamefont {Brixner}},\ }\bibfield  {title} {\bibinfo {title} {Exciton–phonon coupling strength in single-layer mose2 at room temperature},\ }\href {https://doi.org/10.1038/s41467-021-20895-0} {\bibfield  {journal} {\bibinfo  {journal} {Nature Communications}\ }\textbf {\bibinfo {volume} {12}},\ \bibinfo {pages} {954} (\bibinfo {year} {2021})}\BibitemShut {NoStop}%
\bibitem [{\citenamefont {Lu}\ \emph {et~al.}(2016)\citenamefont {Lu}, \citenamefont {Zhang}, \citenamefont {Hwang}, \citenamefont {Ofori-Okai}, \citenamefont {Fleischer},\ and\ \citenamefont {Nelson}}]{Lu2016}%
  \BibitemOpen
  \bibfield  {author} {\bibinfo {author} {\bibfnamefont {J.}~\bibnamefont {Lu}}, \bibinfo {author} {\bibfnamefont {Y.}~\bibnamefont {Zhang}}, \bibinfo {author} {\bibfnamefont {H.~Y.}\ \bibnamefont {Hwang}}, \bibinfo {author} {\bibfnamefont {B.~K.}\ \bibnamefont {Ofori-Okai}}, \bibinfo {author} {\bibfnamefont {S.}~\bibnamefont {Fleischer}},\ and\ \bibinfo {author} {\bibfnamefont {K.~A.}\ \bibnamefont {Nelson}},\ }\bibfield  {title} {\bibinfo {title} {Nonlinear two-dimensional terahertz photon echo and rotational spectroscopy in the gas phase},\ }\href {https://doi.org/10.1073/pnas.1609558113} {\bibfield  {journal} {\bibinfo  {journal} {Proceedings of the National Academy of Sciences}\ }\textbf {\bibinfo {volume} {113}},\ \bibinfo {pages} {11800} (\bibinfo {year} {2016})},\ \Eprint {https://arxiv.org/abs/https://www.pnas.org/doi/pdf/10.1073/pnas.1609558113} {https://www.pnas.org/doi/pdf/10.1073/pnas.1609558113} \BibitemShut {NoStop}%
\bibitem [{\citenamefont {Yu}\ \emph {et~al.}(2019)\citenamefont {Yu}, \citenamefont {Titze}, \citenamefont {Zhu}, \citenamefont {Liu},\ and\ \citenamefont {Li}}]{Yu2019}%
  \BibitemOpen
  \bibfield  {author} {\bibinfo {author} {\bibfnamefont {S.}~\bibnamefont {Yu}}, \bibinfo {author} {\bibfnamefont {M.}~\bibnamefont {Titze}}, \bibinfo {author} {\bibfnamefont {Y.}~\bibnamefont {Zhu}}, \bibinfo {author} {\bibfnamefont {X.}~\bibnamefont {Liu}},\ and\ \bibinfo {author} {\bibfnamefont {H.}~\bibnamefont {Li}},\ }\bibfield  {title} {\bibinfo {title} {Long range dipole-dipole interaction in low-density atomic vapors probed by double-quantum two-dimensional coherent spectroscopy},\ }\href {https://doi.org/10.1364/OE.27.028891} {\bibfield  {journal} {\bibinfo  {journal} {Opt. Express}\ }\textbf {\bibinfo {volume} {27}},\ \bibinfo {pages} {28891} (\bibinfo {year} {2019})}\BibitemShut {NoStop}%
\bibitem [{\citenamefont {Tian}\ \emph {et~al.}(2003)\citenamefont {Tian}, \citenamefont {Keusters}, \citenamefont {Suzaki},\ and\ \citenamefont {Warren}}]{Tian2003}%
  \BibitemOpen
  \bibfield  {author} {\bibinfo {author} {\bibfnamefont {P.}~\bibnamefont {Tian}}, \bibinfo {author} {\bibfnamefont {D.}~\bibnamefont {Keusters}}, \bibinfo {author} {\bibfnamefont {Y.}~\bibnamefont {Suzaki}},\ and\ \bibinfo {author} {\bibfnamefont {W.~S.}\ \bibnamefont {Warren}},\ }\bibfield  {title} {\bibinfo {title} {Femtosecond phase-coherent two-dimensional spectroscopy},\ }\href {https://doi.org/10.1126/science.1083433} {\bibfield  {journal} {\bibinfo  {journal} {Science}\ }\textbf {\bibinfo {volume} {300}},\ \bibinfo {pages} {1553} (\bibinfo {year} {2003})}\BibitemShut {NoStop}%
\bibitem [{\citenamefont {Dai}\ \emph {et~al.}(2012)\citenamefont {Dai}, \citenamefont {Richter}, \citenamefont {Li}, \citenamefont {Bristow}, \citenamefont {Falvo}, \citenamefont {Mukamel},\ and\ \citenamefont {Cundiff}}]{Dai2012}%
  \BibitemOpen
  \bibfield  {author} {\bibinfo {author} {\bibfnamefont {X.}~\bibnamefont {Dai}}, \bibinfo {author} {\bibfnamefont {M.}~\bibnamefont {Richter}}, \bibinfo {author} {\bibfnamefont {H.}~\bibnamefont {Li}}, \bibinfo {author} {\bibfnamefont {A.~D.}\ \bibnamefont {Bristow}}, \bibinfo {author} {\bibfnamefont {C.}~\bibnamefont {Falvo}}, \bibinfo {author} {\bibfnamefont {S.}~\bibnamefont {Mukamel}},\ and\ \bibinfo {author} {\bibfnamefont {S.~T.}\ \bibnamefont {Cundiff}},\ }\bibfield  {title} {\bibinfo {title} {Two-dimensional double-quantum spectra reveal collective resonances in an atomic vapor},\ }\href {https://doi.org/10.1103/PhysRevLett.108.193201} {\bibfield  {journal} {\bibinfo  {journal} {Phys. Rev. Lett.}\ }\textbf {\bibinfo {volume} {108}},\ \bibinfo {pages} {193201} (\bibinfo {year} {2012})}\BibitemShut {NoStop}%
\bibitem [{\citenamefont {Grumstrup}\ \emph {et~al.}(2007)\citenamefont {Grumstrup}, \citenamefont {Shim}, \citenamefont {Montgomery}, \citenamefont {Damrauer},\ and\ \citenamefont {Zanni}}]{Grumstrup2007}%
  \BibitemOpen
  \bibfield  {author} {\bibinfo {author} {\bibfnamefont {E.~M.}\ \bibnamefont {Grumstrup}}, \bibinfo {author} {\bibfnamefont {S.-H.}\ \bibnamefont {Shim}}, \bibinfo {author} {\bibfnamefont {M.~A.}\ \bibnamefont {Montgomery}}, \bibinfo {author} {\bibfnamefont {N.~H.}\ \bibnamefont {Damrauer}},\ and\ \bibinfo {author} {\bibfnamefont {M.~T.}\ \bibnamefont {Zanni}},\ }\bibfield  {title} {\bibinfo {title} {Facile collection of two-dimensional electronic spectra using femtosecond pulse-shaping technology},\ }\href {https://doi.org/10.1364/OE.15.016681} {\bibfield  {journal} {\bibinfo  {journal} {Opt. Express}\ }\textbf {\bibinfo {volume} {15}},\ \bibinfo {pages} {16681} (\bibinfo {year} {2007})}\BibitemShut {NoStop}%
\bibitem [{\citenamefont {Shim}\ and\ \citenamefont {Zanni}(2009)}]{Shim2009}%
  \BibitemOpen
  \bibfield  {author} {\bibinfo {author} {\bibfnamefont {S.-H.}\ \bibnamefont {Shim}}\ and\ \bibinfo {author} {\bibfnamefont {M.~T.}\ \bibnamefont {Zanni}},\ }\bibfield  {title} {\bibinfo {title} {How to turn your pump-probe instrument into a multidimensional spectrometer: 2d ir and vis spectroscopiesvia pulse shaping},\ }\href {https://doi.org/10.1039/B813817F} {\bibfield  {journal} {\bibinfo  {journal} {Phys. Chem. Chem. Phys.}\ }\textbf {\bibinfo {volume} {11}},\ \bibinfo {pages} {748} (\bibinfo {year} {2009})}\BibitemShut {NoStop}%
\bibitem [{\citenamefont {Tekavec}\ \emph {et~al.}(2009)\citenamefont {Tekavec}, \citenamefont {Myers}, \citenamefont {Lewis},\ and\ \citenamefont {Ogilvie}}]{Tekavec2009}%
  \BibitemOpen
  \bibfield  {author} {\bibinfo {author} {\bibfnamefont {P.~F.}\ \bibnamefont {Tekavec}}, \bibinfo {author} {\bibfnamefont {J.~A.}\ \bibnamefont {Myers}}, \bibinfo {author} {\bibfnamefont {K.~L.~M.}\ \bibnamefont {Lewis}},\ and\ \bibinfo {author} {\bibfnamefont {J.~P.}\ \bibnamefont {Ogilvie}},\ }\bibfield  {title} {\bibinfo {title} {Two-dimensional electronic spectroscopy with a continuum probe},\ }\href {https://doi.org/10.1364/OL.34.001390} {\bibfield  {journal} {\bibinfo  {journal} {Opt. Lett.}\ }\textbf {\bibinfo {volume} {34}},\ \bibinfo {pages} {1390} (\bibinfo {year} {2009})}\BibitemShut {NoStop}%
\bibitem [{\citenamefont {Pollard}\ \emph {et~al.}(1989)\citenamefont {Pollard}, \citenamefont {Cruz}, \citenamefont {Shank},\ and\ \citenamefont {Mathies}}]{Pollard1989}%
  \BibitemOpen
  \bibfield  {author} {\bibinfo {author} {\bibfnamefont {W.~T.}\ \bibnamefont {Pollard}}, \bibinfo {author} {\bibfnamefont {C.~H.~B.}\ \bibnamefont {Cruz}}, \bibinfo {author} {\bibfnamefont {C.~V.}\ \bibnamefont {Shank}},\ and\ \bibinfo {author} {\bibfnamefont {R.~A.}\ \bibnamefont {Mathies}},\ }\bibfield  {title} {\bibinfo {title} {Direct observation of the excited‐state cis–trans photoisomerization of bacteriorhodopsin: Multilevel line shape theory for femtosecond dynamic hole burning and its application},\ }\href {https://doi.org/10.1063/1.456658} {\bibfield  {journal} {\bibinfo  {journal} {The Journal of Chemical Physics}\ }\textbf {\bibinfo {volume} {90}},\ \bibinfo {pages} {199} (\bibinfo {year} {1989})}\BibitemShut {NoStop}%
\bibitem [{\citenamefont {Yan}\ and\ \citenamefont {Mukamel}(1990)}]{Yan1990}%
  \BibitemOpen
  \bibfield  {author} {\bibinfo {author} {\bibfnamefont {Y.~J.}\ \bibnamefont {Yan}}\ and\ \bibinfo {author} {\bibfnamefont {S.}~\bibnamefont {Mukamel}},\ }\bibfield  {title} {\bibinfo {title} {Femtosecond pump-probe spectroscopy of polyatomic molecules in condensed phases},\ }\href {https://doi.org/10.1103/PhysRevA.41.6485} {\bibfield  {journal} {\bibinfo  {journal} {Phys. Rev. A}\ }\textbf {\bibinfo {volume} {41}},\ \bibinfo {pages} {6485} (\bibinfo {year} {1990})}\BibitemShut {NoStop}%
\bibitem [{\citenamefont {Berera}\ \emph {et~al.}(2009)\citenamefont {Berera}, \citenamefont {van Grondelle},\ and\ \citenamefont {Kennis}}]{Berera2009}%
  \BibitemOpen
  \bibfield  {author} {\bibinfo {author} {\bibfnamefont {R.}~\bibnamefont {Berera}}, \bibinfo {author} {\bibfnamefont {R.}~\bibnamefont {van Grondelle}},\ and\ \bibinfo {author} {\bibfnamefont {J.~T.~M.}\ \bibnamefont {Kennis}},\ }\bibfield  {title} {\bibinfo {title} {Ultrafast transient absorption spectroscopy: principles and application to photosynthetic systems},\ }\href {https://doi.org/10.1007/s11120-009-9454-y} {\bibfield  {journal} {\bibinfo  {journal} {Photosynthesis Research}\ }\textbf {\bibinfo {volume} {101}},\ \bibinfo {pages} {105} (\bibinfo {year} {2009})}\BibitemShut {NoStop}%
\bibitem [{\citenamefont {Ruckebusch}\ \emph {et~al.}(2012)\citenamefont {Ruckebusch}, \citenamefont {Sliwa}, \citenamefont {Pernot}, \citenamefont {{de Juan}},\ and\ \citenamefont {Tauler}}]{Ruckebusch2012}%
  \BibitemOpen
  \bibfield  {author} {\bibinfo {author} {\bibfnamefont {C.}~\bibnamefont {Ruckebusch}}, \bibinfo {author} {\bibfnamefont {M.}~\bibnamefont {Sliwa}}, \bibinfo {author} {\bibfnamefont {P.}~\bibnamefont {Pernot}}, \bibinfo {author} {\bibfnamefont {A.}~\bibnamefont {{de Juan}}},\ and\ \bibinfo {author} {\bibfnamefont {R.}~\bibnamefont {Tauler}},\ }\bibfield  {title} {\bibinfo {title} {Comprehensive data analysis of femtosecond transient absorption spectra: A review},\ }\href {https://doi.org/https://doi.org/10.1016/j.jphotochemrev.2011.10.002} {\bibfield  {journal} {\bibinfo  {journal} {Journal of Photochemistry and Photobiology C: Photochemistry Reviews}\ }\textbf {\bibinfo {volume} {13}},\ \bibinfo {pages} {1} (\bibinfo {year} {2012})}\BibitemShut {NoStop}%
\bibitem [{\citenamefont {Yue}\ \emph {et~al.}(2023)\citenamefont {Yue}, \citenamefont {Zhou}, \citenamefont {Su}, \citenamefont {Tian}, \citenamefont {Ran},\ and\ \citenamefont {Zhang}}]{Yue2023}%
  \BibitemOpen
  \bibfield  {author} {\bibinfo {author} {\bibfnamefont {J.}~\bibnamefont {Yue}}, \bibinfo {author} {\bibfnamefont {L.}~\bibnamefont {Zhou}}, \bibinfo {author} {\bibfnamefont {P.}~\bibnamefont {Su}}, \bibinfo {author} {\bibfnamefont {L.}~\bibnamefont {Tian}}, \bibinfo {author} {\bibfnamefont {G.}~\bibnamefont {Ran}},\ and\ \bibinfo {author} {\bibfnamefont {W.}~\bibnamefont {Zhang}},\ }\bibfield  {title} {\bibinfo {title} {Complete elimination of pump scattering in transient absorption spectroscopy using phase and amplitude modulation},\ }\href {https://doi.org/https://doi.org/10.1016/j.chemphys.2023.111846} {\bibfield  {journal} {\bibinfo  {journal} {Chemical Physics}\ }\textbf {\bibinfo {volume} {568}},\ \bibinfo {pages} {111846} (\bibinfo {year} {2023})}\BibitemShut {NoStop}%
\bibitem [{\citenamefont {Middleton}\ \emph {et~al.}(2010)\citenamefont {Middleton}, \citenamefont {Woys}, \citenamefont {Mukherjee},\ and\ \citenamefont {Zanni}}]{Chris2010}%
  \BibitemOpen
  \bibfield  {author} {\bibinfo {author} {\bibfnamefont {C.~T.}\ \bibnamefont {Middleton}}, \bibinfo {author} {\bibfnamefont {A.~M.}\ \bibnamefont {Woys}}, \bibinfo {author} {\bibfnamefont {S.~S.}\ \bibnamefont {Mukherjee}},\ and\ \bibinfo {author} {\bibfnamefont {M.~T.}\ \bibnamefont {Zanni}},\ }\bibfield  {title} {\bibinfo {title} {Residue-specific structural kinetics of proteins through the union of isotope labeling, mid-ir pulse shaping, and coherent 2d ir spectroscopy},\ }\href {https://doi.org/https://doi.org/10.1016/j.ymeth.2010.05.002} {\bibfield  {journal} {\bibinfo  {journal} {Methods}\ }\textbf {\bibinfo {volume} {52}},\ \bibinfo {pages} {12} (\bibinfo {year} {2010})},\ \bibinfo {note} {protein Folding}\BibitemShut {NoStop}%
\bibitem [{\citenamefont {Shim}\ \emph {et~al.}(2006)\citenamefont {Shim}, \citenamefont {Strasfeld},\ and\ \citenamefont {Zanni}}]{Shim2006}%
  \BibitemOpen
  \bibfield  {author} {\bibinfo {author} {\bibfnamefont {S.-H.}\ \bibnamefont {Shim}}, \bibinfo {author} {\bibfnamefont {D.~B.}\ \bibnamefont {Strasfeld}},\ and\ \bibinfo {author} {\bibfnamefont {M.~T.}\ \bibnamefont {Zanni}},\ }\bibfield  {title} {\bibinfo {title} {Generation and characterization of phase and amplitude shaped femtosecond mid-ir pulses},\ }\href {https://doi.org/10.1364/OE.14.013120} {\bibfield  {journal} {\bibinfo  {journal} {Opt. Express}\ }\textbf {\bibinfo {volume} {14}},\ \bibinfo {pages} {13120} (\bibinfo {year} {2006})}\BibitemShut {NoStop}%
\bibitem [{\citenamefont {Myers}\ \emph {et~al.}(2008)\citenamefont {Myers}, \citenamefont {Lewis}, \citenamefont {Tekavec},\ and\ \citenamefont {Ogilvie}}]{Myers2008}%
  \BibitemOpen
  \bibfield  {author} {\bibinfo {author} {\bibfnamefont {J.~A.}\ \bibnamefont {Myers}}, \bibinfo {author} {\bibfnamefont {K.~L.~M.}\ \bibnamefont {Lewis}}, \bibinfo {author} {\bibfnamefont {P.~F.}\ \bibnamefont {Tekavec}},\ and\ \bibinfo {author} {\bibfnamefont {J.~P.}\ \bibnamefont {Ogilvie}},\ }\bibfield  {title} {\bibinfo {title} {Two-color two-dimensional fourier transform electronic spectroscopy with a pulse-shaper},\ }\href {https://doi.org/10.1364/OE.16.017420} {\bibfield  {journal} {\bibinfo  {journal} {Opt. Express}\ }\textbf {\bibinfo {volume} {16}},\ \bibinfo {pages} {17420} (\bibinfo {year} {2008})}\BibitemShut {NoStop}%
\bibitem [{\citenamefont {Song}\ \emph {et~al.}(2021)\citenamefont {Song}, \citenamefont {Sechrist}, \citenamefont {Nguyen}, \citenamefont {Johnson}, \citenamefont {Abramavicius}, \citenamefont {Redding},\ and\ \citenamefont {Ogilvie}}]{Song2021}%
  \BibitemOpen
  \bibfield  {author} {\bibinfo {author} {\bibfnamefont {Y.}~\bibnamefont {Song}}, \bibinfo {author} {\bibfnamefont {R.}~\bibnamefont {Sechrist}}, \bibinfo {author} {\bibfnamefont {H.~H.}\ \bibnamefont {Nguyen}}, \bibinfo {author} {\bibfnamefont {W.}~\bibnamefont {Johnson}}, \bibinfo {author} {\bibfnamefont {D.}~\bibnamefont {Abramavicius}}, \bibinfo {author} {\bibfnamefont {K.~E.}\ \bibnamefont {Redding}},\ and\ \bibinfo {author} {\bibfnamefont {J.~P.}\ \bibnamefont {Ogilvie}},\ }\bibfield  {title} {\bibinfo {title} {Excitonic structure and charge separation in the heliobacterial reaction center probed by multispectral multidimensional spectroscopy},\ }\href {https://doi.org/10.1038/s41467-021-23060-9} {\bibfield  {journal} {\bibinfo  {journal} {Nature Communications}\ }\textbf {\bibinfo {volume} {12}},\ \bibinfo {pages} {2801} (\bibinfo {year} {2021})}\BibitemShut {NoStop}%
\bibitem [{\citenamefont {Megerle}\ \emph {et~al.}(2009)\citenamefont {Megerle}, \citenamefont {Pugliesi}, \citenamefont {Schriever}, \citenamefont {Sailer},\ and\ \citenamefont {Riedle}}]{Megerle2009}%
  \BibitemOpen
  \bibfield  {author} {\bibinfo {author} {\bibfnamefont {U.}~\bibnamefont {Megerle}}, \bibinfo {author} {\bibfnamefont {I.}~\bibnamefont {Pugliesi}}, \bibinfo {author} {\bibfnamefont {C.}~\bibnamefont {Schriever}}, \bibinfo {author} {\bibfnamefont {C.~F.}\ \bibnamefont {Sailer}},\ and\ \bibinfo {author} {\bibfnamefont {E.}~\bibnamefont {Riedle}},\ }\bibfield  {title} {\bibinfo {title} {Sub-50 fs broadband absorption spectroscopy with tunable excitation: putting the analysis of ultrafast molecular dynamics on solid ground},\ }\href {https://doi.org/10.1007/s00340-009-3610-0} {\bibfield  {journal} {\bibinfo  {journal} {Applied Physics B}\ }\textbf {\bibinfo {volume} {96}},\ \bibinfo {pages} {215} (\bibinfo {year} {2009})}\BibitemShut {NoStop}%
\bibitem [{\citenamefont {Liebel}\ \emph {et~al.}(2015)\citenamefont {Liebel}, \citenamefont {Schnedermann}, \citenamefont {Wende},\ and\ \citenamefont {Kukura}}]{Liebel2015}%
  \BibitemOpen
  \bibfield  {author} {\bibinfo {author} {\bibfnamefont {M.}~\bibnamefont {Liebel}}, \bibinfo {author} {\bibfnamefont {C.}~\bibnamefont {Schnedermann}}, \bibinfo {author} {\bibfnamefont {T.}~\bibnamefont {Wende}},\ and\ \bibinfo {author} {\bibfnamefont {P.}~\bibnamefont {Kukura}},\ }\bibfield  {title} {\bibinfo {title} {Principles and applications of broadband impulsive vibrational spectroscopy},\ }\href {https://doi.org/10.1021/acs.jpca.5b05948} {\bibfield  {journal} {\bibinfo  {journal} {The Journal of Physical Chemistry A}\ }\textbf {\bibinfo {volume} {119}},\ \bibinfo {pages} {9506} (\bibinfo {year} {2015})},\ \bibinfo {note} {pMID: 26262557}\BibitemShut {NoStop}%
\bibitem [{\citenamefont {Tekavec}\ \emph {et~al.}(2010)\citenamefont {Tekavec}, \citenamefont {Myers}, \citenamefont {Lewis}, \citenamefont {Fuller},\ and\ \citenamefont {Ogilvie}}]{Tekavec2010}%
  \BibitemOpen
  \bibfield  {author} {\bibinfo {author} {\bibfnamefont {P.~F.}\ \bibnamefont {Tekavec}}, \bibinfo {author} {\bibfnamefont {J.~A.}\ \bibnamefont {Myers}}, \bibinfo {author} {\bibfnamefont {K.~L.~M.}\ \bibnamefont {Lewis}}, \bibinfo {author} {\bibfnamefont {F.~D.}\ \bibnamefont {Fuller}},\ and\ \bibinfo {author} {\bibfnamefont {J.~P.}\ \bibnamefont {Ogilvie}},\ }\bibfield  {title} {\bibinfo {title} {Effects of chirp on two-dimensional fourier transform electronic spectra},\ }\href {https://doi.org/10.1364/OE.18.011015} {\bibfield  {journal} {\bibinfo  {journal} {Opt. Express}\ }\textbf {\bibinfo {volume} {18}},\ \bibinfo {pages} {11015} (\bibinfo {year} {2010})}\BibitemShut {NoStop}%
\bibitem [{\citenamefont {Tekavec}\ \emph {et~al.}(2012)\citenamefont {Tekavec}, \citenamefont {Lewis}, \citenamefont {Fuller}, \citenamefont {Myers},\ and\ \citenamefont {Ogilvie}}]{Tekavec2012}%
  \BibitemOpen
  \bibfield  {author} {\bibinfo {author} {\bibfnamefont {P.~A.}\ \bibnamefont {Tekavec}}, \bibinfo {author} {\bibfnamefont {K.~L.~M.}\ \bibnamefont {Lewis}}, \bibinfo {author} {\bibfnamefont {F.~D.}\ \bibnamefont {Fuller}}, \bibinfo {author} {\bibfnamefont {J.~A.}\ \bibnamefont {Myers}},\ and\ \bibinfo {author} {\bibfnamefont {J.~P.}\ \bibnamefont {Ogilvie}},\ }\bibfield  {title} {\bibinfo {title} {Toward broad bandwidth 2-d electronic spectroscopy: Correction of chirp from a continuum probe},\ }\href {https://doi.org/10.1109/JSTQE.2011.2109941} {\bibfield  {journal} {\bibinfo  {journal} {IEEE Journal of Selected Topics in Quantum Electronics}\ }\textbf {\bibinfo {volume} {18}},\ \bibinfo {pages} {210} (\bibinfo {year} {2012})}\BibitemShut {NoStop}%
\bibitem [{\citenamefont {Kovalenko}\ \emph {et~al.}(1999)\citenamefont {Kovalenko}, \citenamefont {Dobryakov}, \citenamefont {Ruthmann},\ and\ \citenamefont {Ernsting}}]{Ernsting1999}%
  \BibitemOpen
  \bibfield  {author} {\bibinfo {author} {\bibfnamefont {S.~A.}\ \bibnamefont {Kovalenko}}, \bibinfo {author} {\bibfnamefont {A.~L.}\ \bibnamefont {Dobryakov}}, \bibinfo {author} {\bibfnamefont {J.}~\bibnamefont {Ruthmann}},\ and\ \bibinfo {author} {\bibfnamefont {N.~P.}\ \bibnamefont {Ernsting}},\ }\bibfield  {title} {\bibinfo {title} {Femtosecond spectroscopy of condensed phases with chirped supercontinuum probing},\ }\href {https://doi.org/10.1103/PhysRevA.59.2369} {\bibfield  {journal} {\bibinfo  {journal} {Phys. Rev. A}\ }\textbf {\bibinfo {volume} {59}},\ \bibinfo {pages} {2369} (\bibinfo {year} {1999})}\BibitemShut {NoStop}%
\bibitem [{\citenamefont {Gardecki}\ \emph {et~al.}(2000)\citenamefont {Gardecki}, \citenamefont {Constantine}, \citenamefont {Zhou},\ and\ \citenamefont {Ziegler}}]{Gardecki2000}%
  \BibitemOpen
  \bibfield  {author} {\bibinfo {author} {\bibfnamefont {J.~A.}\ \bibnamefont {Gardecki}}, \bibinfo {author} {\bibfnamefont {S.}~\bibnamefont {Constantine}}, \bibinfo {author} {\bibfnamefont {Y.}~\bibnamefont {Zhou}},\ and\ \bibinfo {author} {\bibfnamefont {L.~D.}\ \bibnamefont {Ziegler}},\ }\bibfield  {title} {\bibinfo {title} {Optical heterodyne detected spectrograms of ultrafast nonresonant electronic responses},\ }\href {https://doi.org/10.1364/JOSAB.17.000652} {\bibfield  {journal} {\bibinfo  {journal} {J. Opt. Soc. Am. B}\ }\textbf {\bibinfo {volume} {17}},\ \bibinfo {pages} {652} (\bibinfo {year} {2000})}\BibitemShut {NoStop}%
\bibitem [{\citenamefont {Yeremenko}\ \emph {et~al.}(2002)\citenamefont {Yeremenko}, \citenamefont {Baltu\v{s}ka}, \citenamefont {de~Haan}, \citenamefont {Pshenichnikov},\ and\ \citenamefont {Wiersma}}]{Yeremenko2002}%
  \BibitemOpen
  \bibfield  {author} {\bibinfo {author} {\bibfnamefont {S.}~\bibnamefont {Yeremenko}}, \bibinfo {author} {\bibfnamefont {A.}~\bibnamefont {Baltu\v{s}ka}}, \bibinfo {author} {\bibfnamefont {F.}~\bibnamefont {de~Haan}}, \bibinfo {author} {\bibfnamefont {M.~S.}\ \bibnamefont {Pshenichnikov}},\ and\ \bibinfo {author} {\bibfnamefont {D.~A.}\ \bibnamefont {Wiersma}},\ }\bibfield  {title} {\bibinfo {title} {Frequency-resolved pump--probe characterization of femtosecond infrared pulses},\ }\href {https://doi.org/10.1364/OL.27.001171} {\bibfield  {journal} {\bibinfo  {journal} {Opt. Lett.}\ }\textbf {\bibinfo {volume} {27}},\ \bibinfo {pages} {1171} (\bibinfo {year} {2002})}\BibitemShut {NoStop}%
\bibitem [{\citenamefont {Lorenc}\ \emph {et~al.}(2002)\citenamefont {Lorenc}, \citenamefont {Ziolek}, \citenamefont {Naskrecki}, \citenamefont {Karolczak}, \citenamefont {Kubicki},\ and\ \citenamefont {Maciejewski}}]{Lorenc2002}%
  \BibitemOpen
  \bibfield  {author} {\bibinfo {author} {\bibfnamefont {M.}~\bibnamefont {Lorenc}}, \bibinfo {author} {\bibfnamefont {M.}~\bibnamefont {Ziolek}}, \bibinfo {author} {\bibfnamefont {R.}~\bibnamefont {Naskrecki}}, \bibinfo {author} {\bibfnamefont {J.}~\bibnamefont {Karolczak}}, \bibinfo {author} {\bibfnamefont {J.}~\bibnamefont {Kubicki}},\ and\ \bibinfo {author} {\bibfnamefont {A.}~\bibnamefont {Maciejewski}},\ }\bibfield  {title} {\bibinfo {title} {Artifacts in femtosecond transient absorption spectroscopy},\ }\href {https://doi.org/10.1007/s003400100750} {\bibfield  {journal} {\bibinfo  {journal} {Applied Physics B}\ }\textbf {\bibinfo {volume} {74}},\ \bibinfo {pages} {19} (\bibinfo {year} {2002})}\BibitemShut {NoStop}%
\bibitem [{\citenamefont {Ekvall}\ \emph {et~al.}(2000)\citenamefont {Ekvall}, \citenamefont {van~der Meulen}, \citenamefont {Dhollande}, \citenamefont {Berg}, \citenamefont {Pommeret}, \citenamefont {Naskrecki},\ and\ \citenamefont {Mialocq}}]{Ekvall2000}%
  \BibitemOpen
  \bibfield  {author} {\bibinfo {author} {\bibfnamefont {K.}~\bibnamefont {Ekvall}}, \bibinfo {author} {\bibfnamefont {P.}~\bibnamefont {van~der Meulen}}, \bibinfo {author} {\bibfnamefont {C.}~\bibnamefont {Dhollande}}, \bibinfo {author} {\bibfnamefont {L.-E.}\ \bibnamefont {Berg}}, \bibinfo {author} {\bibfnamefont {S.}~\bibnamefont {Pommeret}}, \bibinfo {author} {\bibfnamefont {R.}~\bibnamefont {Naskrecki}},\ and\ \bibinfo {author} {\bibfnamefont {J.-C.}\ \bibnamefont {Mialocq}},\ }\bibfield  {title} {\bibinfo {title} {{Cross phase modulation artifact in liquid phase transient absorption spectroscopy}},\ }\href {https://doi.org/10.1063/1.372185} {\bibfield  {journal} {\bibinfo  {journal} {Journal of Applied Physics}\ }\textbf {\bibinfo {volume} {87}},\ \bibinfo {pages} {2340} (\bibinfo {year} {2000})}\BibitemShut {NoStop}%
\bibitem [{\citenamefont {Park}(2015)}]{Park2015}%
  \BibitemOpen
  \bibfield  {author} {\bibinfo {author} {\bibfnamefont {S.~D.}\ \bibnamefont {Park}},\ }\emph {\bibinfo {title} {Femtosecond and Two-Dimensional Spectroscopy of Lead Chalcogenide Quantum Dots}},\ \href {https://scholar.colorado.edu/concern/graduate_thesis_or_dissertations/hd76s016r} {\bibinfo {type} {Dissertation}} (\bibinfo {year} {2015})\BibitemShut {NoStop}%
\bibitem [{\citenamefont {Yue}\ \emph {et~al.}(2015)\citenamefont {Yue}, \citenamefont {Wang}, \citenamefont {He}, \citenamefont {Zhu},\ and\ \citenamefont {Weng}}]{Yue2015}%
  \BibitemOpen
  \bibfield  {author} {\bibinfo {author} {\bibfnamefont {S.}~\bibnamefont {Yue}}, \bibinfo {author} {\bibfnamefont {Z.}~\bibnamefont {Wang}}, \bibinfo {author} {\bibfnamefont {X.-c.}\ \bibnamefont {He}}, \bibinfo {author} {\bibfnamefont {G.-b.}\ \bibnamefont {Zhu}},\ and\ \bibinfo {author} {\bibfnamefont {Y.-x.}\ \bibnamefont {Weng}},\ }\bibfield  {title} {\bibinfo {title} {{Construction of the Apparatus for Two Dimensional Electronic Spectroscopy and Characterization of the Instrument†}},\ }\href {https://doi.org/10.1063/1674-0068/28/cjcp1506136} {\bibfield  {journal} {\bibinfo  {journal} {Chinese Journal of Chemical Physics}\ }\textbf {\bibinfo {volume} {28}},\ \bibinfo {pages} {509} (\bibinfo {year} {2015})}\BibitemShut {NoStop}%
\bibitem [{\citenamefont {Zhang}\ and\ \citenamefont {Dong}(2022)}]{Zhang2022}%
  \BibitemOpen
  \bibfield  {author} {\bibinfo {author} {\bibfnamefont {X.}~\bibnamefont {Zhang}}\ and\ \bibinfo {author} {\bibfnamefont {H.}~\bibnamefont {Dong}},\ }\bibfield  {title} {\bibinfo {title} {Nonperturbative approach to the nonlinear photon echo of a v-type system},\ }\href {https://doi.org/10.1103/PhysRevA.106.043516} {\bibfield  {journal} {\bibinfo  {journal} {Phys. Rev. A}\ }\textbf {\bibinfo {volume} {106}},\ \bibinfo {pages} {043516} (\bibinfo {year} {2022})}\BibitemShut {NoStop}%
\bibitem [{\citenamefont {Oudar}(1983)}]{Oudar1983}%
  \BibitemOpen
  \bibfield  {author} {\bibinfo {author} {\bibfnamefont {J.-L.}\ \bibnamefont {Oudar}},\ }\bibfield  {title} {\bibinfo {title} {Coherent phenomena involved in the time-resolved optical kerr effect},\ }\href {https://doi.org/10.1109/JQE.1983.1071918} {\bibfield  {journal} {\bibinfo  {journal} {IEEE Journal of Quantum Electronics}\ }\textbf {\bibinfo {volume} {19}},\ \bibinfo {pages} {713} (\bibinfo {year} {1983})}\BibitemShut {NoStop}%
\bibitem [{\citenamefont {Hellwarth}\ \emph {et~al.}(1975)\citenamefont {Hellwarth}, \citenamefont {Cherlow},\ and\ \citenamefont {Yang}}]{Hellwarth1975}%
  \BibitemOpen
  \bibfield  {author} {\bibinfo {author} {\bibfnamefont {R.}~\bibnamefont {Hellwarth}}, \bibinfo {author} {\bibfnamefont {J.}~\bibnamefont {Cherlow}},\ and\ \bibinfo {author} {\bibfnamefont {T.-T.}\ \bibnamefont {Yang}},\ }\bibfield  {title} {\bibinfo {title} {Origin and frequency dependence of nonlinear optical susceptibilities of glasses},\ }\href {https://doi.org/10.1103/PhysRevB.11.964} {\bibfield  {journal} {\bibinfo  {journal} {Phys. Rev. B}\ }\textbf {\bibinfo {volume} {11}},\ \bibinfo {pages} {964} (\bibinfo {year} {1975})}\BibitemShut {NoStop}%
\bibitem [{\citenamefont {Alfano}\ and\ \citenamefont {Shapiro}(1970)}]{Alfano1970}%
  \BibitemOpen
  \bibfield  {author} {\bibinfo {author} {\bibfnamefont {R.~R.}\ \bibnamefont {Alfano}}\ and\ \bibinfo {author} {\bibfnamefont {S.~L.}\ \bibnamefont {Shapiro}},\ }\bibfield  {title} {\bibinfo {title} {Observation of self-phase modulation and small-scale filaments in crystals and glasses},\ }\href {https://doi.org/10.1103/PhysRevLett.24.592} {\bibfield  {journal} {\bibinfo  {journal} {Phys. Rev. Lett.}\ }\textbf {\bibinfo {volume} {24}},\ \bibinfo {pages} {592} (\bibinfo {year} {1970})}\BibitemShut {NoStop}%
\end{thebibliography}%

\end{document}